Chromosomal rearrangements as a source of new gene formation in *Drosophila yakuba*

Nicholas B. Stewart[1,2*] and Rebekah L. Rogers[1]

1. Department of Bioinformatics and Genomics, University of North Carolina, Charlotte, NC
2. Department of Biological Sciences, Ft Hays State University, Ft Hays, KS

Corresponding Author*: Rebekah L. Rogers
Department of Bioinformatics and Genomics, University of North Carolina, Charlotte, NC
9201 University City Blvd, Charlotte, NC 28205
Phone: 704-687-1321
Email: rebekah.rogers@uncc.edu

**Abstract**

The origins of new genes are among the most fundamental questions in evolutionary biology. Our understanding of the ways that new genetic material appears and how that genetic material shapes population variation remains incomplete. *De novo* genes and duplicate genes are a key source of new genetic material on which selection acts. To better understand the origins of these new gene sequences, we explored the ways that structural variation might alter expression patterns and form novel transcripts. We provide evidence that chromosomal rearrangements are a source of novel genetic variation that facilitates the formation of *de novo* exons in *Drosophila*. We identify 51 cases of *de novo* exon formation created by chromosomal rearrangements in 14 strains of *D. yakuba*. These new genes inherit transcription start signals and open reading frames when the 5' end of existing genes are combined with previously untranscribed regions. Such new genes would appear with novel peptide sequences, without the necessity for secondary transitions from non-coding RNA to protein. This mechanism of new peptide formations contrasts with canonical theory of *de novo* gene progression requiring non-coding intermediaries that must acquire new mutations prior to loss via pseudogenization. Hence, these mutations offer a means to *de novo* gene creation and protein sequence formation in a single mutational step, answering a long standing open question concerning new gene formation. We further identify gene expression changes to 134 existing genes, indicating that these mutations can alter gene regulation. Population variability for chromosomal rearrangements is considerable, with 2368 rearrangements observed across 14 inbred lines. More rearrangements were identified on the X chromosome than any of the autosomes, suggesting the X is more susceptible to chromosome alterations. Together, these results suggest that chromosomal rearrangements are a source of variation in populations that is likely to be important to explain genetic and therefore phenotypic diversity.

## Introduction

Understanding the origins of new genes is essential for a complete description of evolutionary processes. Mutation generates the raw genetic material that can contribute to phenotypic diversity in natural populations. Without this new genetic material, selection cannot produce change. When new genes are required for adaptation to new and changing environments, where do they come from? How do they arise? Proposed sources of new genetic material include duplicate genes (1, 2), chimeric genes (3-5), *de novo* genes (6-9), and domesticated transposable elements (10). Deep sequencing of genomes has made it trivial to identify single nucleotide polymorphisms (SNPs) in population genetic data (11, 12). In contrast, structural variants and duplications remain understudied, in part because they are more difficult to identify in sequence data. With improvements in throughput and quality of next generation sequencing, we can begin to explore the full effects of these complex mutations in nature.

Chromosomal rearrangements contribute to genomic divergence across species. While organisms exhibit striking similarity in genome content, genome organization becomes scrambled over time breaking syntenic blocks (13-15). Gene movement due to chromosomal rearrangements is known to influence gene expression in primates (16). Such natural variation from genomic neighborhood is similar to positional effects observed in transgenic constructs (17). Yet, the implications of these mutations go far beyond quantitative changes in mRNA levels. Mutations that copy and shuffle pieces of DNA can produce new gene sequences. They have the potential to form whole gene duplications, chimeric genes or alternative gene constructs. The full spectrum of new gene creation from these mutations is not fully explored.

Recent work has identified new gene formation through *de novo* exon creation when duplicated segments do not respect gene boundaries (18). These new genes may ascribe a genetic cause to patterns that mimic *de novo* gene creation. Similar cases of new gene formation were observed when inversions modified gene sequences at breakpoints (19). However, whole genome studies of rearrangements and new gene formation in natural populations have been lacking. We hypothesize that chromosomal rearrangements may form similar new gene structures when they copy or move pieces of DNA around the genome.

*Drosophila* remain an excellent model system for genomic analysis. Their genomes are compact with little repetitive DNA content, and easily sequenced (20). Among the *Drosophila*, *D. yakuba* houses an unusually large number of chromosomal rearrangements based on reference strain comparisons (20). Yet, the complexity of population variation for chromosomal rearrangements that might give rise to the observed divergence remains unseen. Here, we use whole genome population resequencing data with paired-end Illumina reads (21) to identify genome structure changes that are segregating in natural populations of *D. yakuba*. Pairing these mutation scans with high throughput gene expression data (18), we identify regulatory changes that are produced via chromosomal rearrangements. Across 14 inbred lines of *D. yakuba* we used abnormal paired-end read mapping to identify chromosomal rearrangements between and within chromosomes. These mutations may be caused by ectopic recombination, TE movement, inversion formation, template switching during DNA synthesis, ectopic DNA repair, or retrogene formation. These mutations all have the unifying feature that they copy or move DNA from one location to another. We describe the number, types, and locations of rearrangements in population sequence data. Using RNA sequence data from these lines we identified incidences where rearrangements may be creating *de novo* exons and chimeric constructs that create new exons in genomic regions previously devoid of expression. These results suggest that chromosomal rearrangements are a key source of new gene creation that reshapes genome content and organization in nature.

## Results

### Abundance of chromosomal rearrangements

We used abnormally mapping Illumina paired-end sequence reads to survey chromosomal rearrangements in a previously sequenced population of *Drosophila yakuba* (21). We remapped paired-end reads to the reference genome r.1.05 of *Drosophila yakuba* (20) and *Wolbachia* endoparasite sequence NC_002978.6. After removing PCR duplicates, the average depth of coverage of each line varied between 12X to 93X coverage (Table 1). We identified regions that are supported by at least 4 independent read-pairs that map at least 1 Mb away from each other on the same chromosome or on separate chromosomes, similar to methods previously implemented in human genetics (22) (see Figure 1). We identified 2368 total rearrangements among the 14 lines of *Drosophila yakuba*: 1697 rearrangements between chromosomes and 671 within chromosomes. These rearrangements lie within 1kb of 1202 genes.

### Rearrangements as facilitators of new gene formation

Structural variants and duplications can form new exon sequences in regions that were previously untranscribed (18, 22). These new transcripts appear when gene fragments carrying promoter sequences can drive transcription of new exons with new open reading frames. Formation of such chimeric constructs with *de novo* exons can offer a source of new transcripts within the genome. Like most cases of new gene formation, we expect many of these new variants to be transient rather than stably incorporated into the genome (23-25). However, as a substrate of genetic novelty, they may occasionally contribute to adaptive changes and phenotypic variation in nature. New genes often appear with expression in the germline (7, 26, 27). To explore such cases of new open reading frames in the tissues most likely to be affected, we compared rearrangement calls with previously published RNASeq data from testes and accessory glands, male carcasses, ovaries, and female carcasses (18).

To identify new chimeric transcripts formed at rearrangement breakpoints, we used Tophat fusion search (28) to find split reads and abnormally mapping read pairs in RNAseq data (Figure 2). These methods were developed to identify trans-spliced transcripts, but should also identify support for chimeric transcripts produced when genome sequences have been rearranged. We matched structure calls with Tophat fusion (28) calls that are within 1 kb of both sides of rearrangements called in genomic DNA sequences. We used RNASeq read depth to further infer structure of *de novo* transcripts created by chromosome rearrangements (Figure 3) (Figure 4) (S1 Figure) (S2 Figure).

In 14 inbred lines, we isolated 51 putative new genes created by rearrangements. A total of 43 genome structure calls match RNASeq fusion calls that indicate expression in testes. Among such new genes, 32 also show expression in male somatic tissue, while 10 of them are expressed exclusively in the testes. A total of 42 fusion genes were found in male somatic tissue, 33 of which were shared in testes and 9 expressed exclusively in the male somatic tissue. A total of 40 out of 51 transcripts incorporate the start codon of a pre-existing gene (Supporting Information). These data suggest that the majority of new genes that form do so by copying or shuffling 5' promoters and translation signals of genes to drive transcription in new regions. Some 37% (19/51) of the possible new genes identified are singletons (S3 Figure). This pattern is consistent with most of the rearrangements and new gene formation being relatively young and possibly detrimental on average. Additionally, 29% (15/51) of the rearrangements may be ancestral, whereas the reference had a rearrangement that modified a preexisting gene.

We observe more new RNA fusion calls at loci with structural rearrangements in testes and male carcasses than ovaries and female carcasses (Figure 5). Testes and male carcass RNA were sequenced using paired-end sequencing while ovaries and female carcass RNA were sequenced using single end sequencing methods. The use of single end data from previously published work reduces the ability to identify fusion transcripts through split read mapping in females (Supporting Information). We do not see differences between germline and somatic tissue in males (ANOVA, $F(1,13)=0.04$, $P>0.8$). In females there is a difference between gametic and somatic tissues (ANOVA, $F(1,13)=4.379$, $P<0.05$) though samples sizes are small. In total, we found new transcripts produced by 19 rearrangements expressed in females: 14 expressed in ovaries and 14 expressed female carcass. These data do not allow us to identify every rearrangement in the population, largely due to limits in sequencing coverage. Estimates reported here are conservative, representing the minimum number of instances of new gene formation.

We confirmed cases of new gene formation using reference free transcriptome assembly program Trinity v.2.4.0. These transcriptomes were assembled then aligned using BLASTn to the *D. yakuba* reference. A total of 38 *de novo* transcripts (75%) were confirmed by Trinity reference-free transcriptome assembly (29). However, 5 out of the 13 transcripts that could not be confirmed appear in regions that exhibit multiple rearrangements thus making it difficult to confirm with Trinity. Also, Trinity confirmation rates may be reduced when very small exons fail to align to the reference genome at stringent thresholds (S2 Figure). Thus, 75% represents a minimum confirmation rate.

Of the 38 transcripts that were confirmed with Trinity, start codons were located before the breakpoint in 34 (89%) of the transcripts. This would suggest that most of the putative new genes identified are chimeric. Hence, most rearrangements appear to incorporate the 5' UTR and start codon of a pre-existing transcript, thereby forming *de novo* exons. These chimeric constructs with *de novo* exons are a source of new transcript formation that can contribute to variation for gene content in natural populations. As with most new gene formation, we expect many of these new genes to be transient (23-25). However, some small subset may form genetic variation that may be useful for adaptive change.

**Regulatory changes and chromosomal rearrangements**

Chromosomal rearrangements can cause expression changes even when exon sequences remain unmodified (16). To explore such regulatory changes, we used Cuffdiff from the Tophat/Cufflinks gene expression testing suite (30, 31) to identify genes that have significant change of expression compared to the reference strain. We identified 134 genes within 1kb of a rearrangement that had significant expression differences in at least one tissue compared to the reference strain. These include 41 genes in the testes, 51 in male carcass, 50 in ovaries, and 36 in female carcass that show differential expression associated with rearrangements. Most changes in gene expression associated with chromosomal rearrangements produce decreased expression (S2 Table). Such gene expression changes have the potential to induce phenotypic changes in natural populations.

**Population diversity for chromosomal rearrangements**

The number of rearrangements identified per line varies from 96 to 455 total rearrangements in a single strain (Table 1). Low coverage PacBio long molecule data confirmed 80-97% of mutations per strain suggesting a low false positive rate. Sequencing coverage has a strong effect on false negative rates and confirmation rates (see Supporting Information). Mutations were polarized against the ancestral state using a BLASTn against *D. erecta*. We identified 112 (4.7%) rearrangements that represent new mutations in the *D. yakuba* reference. A total of 54 out of 2368 rearrangements could not be polarized using the existing reference assembly (Supporting Information). These are excluded from the site frequency spectrum below. The SFS corrected for false negatives shows that the majority of variants are singletons (Figure 6). This is expected if most of the rearrangements are young and/or have negative fitness.

If rearrangements create novel gene structures or alter gene expression, they may cause phenotypic effects that are subject to natural selection. We wondered whether signatures of selective sweeps might be observed at loci containing rearrangements. Sweep like signals of negative Tajima's *D* representing highly skewed SFS for the region are not overrepresented among rearrangements (Supporting Information). Some rearrangements showed Tajima's *D* in the bottom 5% of all windows in spite of being singleton variants. These likely represent rearrangements that appeared after the incidence of the sweep. These low frequency variants are not candidates for adaptive changes. However, we observe 10 rearrangements found in at least 75% of lines that are also associated with sweep-like signals (Supporting Information). Hence, rearrangements do not appear to be selectively favored as a class, though some individual rearrangements could be adaptive.

**Association with transposable elements**

Transposable elements are known to facilitate chromosomal rearrangements in *Drosophila (32)*. They move DNA from one location to another, sometimes creating duplications. TEs can also facilitate ectopic recombination as repetitive sequence mis-pairs during meiosis or mitosis (32). We compared our rearrangement calls (corrected for false negatives) with TE calls in these lines described previously (21, 33). We found that 694 rearrangement calls have a TE within 1 kb to one of the sites of the rearrangement and 215 rearrangements have a TE within 1 kb of both sites of the rearrangement. Overall 23.7% (1124/4736) of the rearrangement sites lie within 1 kb of a TE. We found 349 (14.7%) rearrangements have reads that overlap directly with at TE. These rearrangements are confirmed at 86.5-100% in PacBio data, similar the genome wide average. Transposable element content in *Drosophila* is limited compared with other animals. Only 5.5% of the reference genome is composed of TEs (20), though TEs may accumulate in poorly assembled heterochromatic regions. Yet, these selfish genetic elements appear to contribute significantly to polymorphic changes in genome content and organization.

**Genomic distribution of chromosome rearrangement breakpoints**

Previous work has noted that the X chromosome is a source of newly transposed transcripts, and sex chromosomes are prone to rearrangements due to repetitive content. An excess of genome structure variants involving the sex chromosomes would leave signals of at least 1 breakpoint lying on the X. We identified the distribution of rearrangement breakpoints within each chromosome arm (Figure 7). We standardized the abundance of rearrangements by the length of chromosome arm. We excluded the 4$^{th}$ chromosome (Muller element F) from our analysis. For rearrangements within a chromosome we excluded abnormally mapping read-pairs less than 1 Mb apart. The chromosome arms have unequal abundance of rearrangement breakpoints per base pair (MANOVA, $F(4, 52)=12.35$, $P<10^{-11}$) (Figure 7). Within-chromosome rearrangements account for 28% of rearrangements, roughly proportional to the amount of the genome housed in a major chromosome arm. These results suggest that the landing place for rearrangements is not biased towards or away from the same chromosome arm.

The X chromosome has significantly more rearrangement breakpoints per base pair than the 4 major autosomes arms ($P<10^{-6}$ for each pairwise comparison; S3 Table). The data reveal that 3R has a reduced number of rearrangement sites compared to the X ($P<10^{-7}$; Sup Table 3), 2L ($P$<0.05; S3 Table), and 2R ($P$<0.002; S3 Table). The excess of rearrangements on the X is consistent with previous findings of an abundance of tandem duplications located on the X in *D. yakuba* (21). The X chromosome has more repetitive regions that are more susceptible to ectopic recombination (34, 35**).** When we distinguish rearrangements based on whether they move DNA across different chromosome arms or affect distant regions on single chromosome arms, the X is still overrepresented (Supplementary Information Text, S6 Figure, S4-S5 Table).

We identified 4 'hotspots' of TE movement that had over 30 rearrangement breakpoints in a 5kb span across 14 lines (Figure 8, S7 Figure). One of these hotspots on 2R lies adjacent to a known inversion breakpoint that is expected to suppress recombination. These hotspots contain sequences matching TE families, consistent with TE proliferation (S1 Text). Most rearrangements at hotspots are singleton variants, and each line has fewer than 10 rearrangements. These results suggest recurrent, independent mutations affecting specific regions of the genome.

**Complex variation**

Many of our rearrangements are found in clustered pairs, most likely reflecting two breakpoints of an insertion. If the insertion is large enough our methods will separate them into two different rearrangement calls. In other cases where the insertion is small and roughly equal to the read length, our methods make only a single rearrangement call. Some rearrangements appear to be more complex than a simple rearrangement of one sequence transferring to a new location, a challenge for paired-end read mapping. Among the data, one example stands out as an unusually labile region. Chromosome 2R houses a 2.5kb region (2R:7003000-7005000) that has up to 5 rearrangements with a 7kb region 2MB up stream (2R:9895000-9902000) (Figure 8). All lines have at least one rearrangement in this region, and 13/14 of the lines have supporting RNASeq data. This region may have undergone a recent selective sweep (2R:7003000-7005000; Tajima's $D$= -2.1364) (2R:9895000-9902000; Tajima's $D$= -1.8177). Due to the multiple rearrangements affecting this single region, it is difficult to localize changes to transcripts and gene expression using Illumina data. This region was identified previously as containing an inversion (34). The multiple rearrangement calls suggest that the inversion possibly is accompanied by multiple duplication events which is also consistent with targeted analysis of this region (34). Regions such as this one represents dynamic genome sequence with multiple changes in a short time. Further research of complex regions, especially with emerging long read technology, may allow for a better understanding of how genes are affected by multiple relative recent changes (35). Such future work may provide an even more complete account for the consequences of chromosomal rearrangements on gene expression and new gene formation.

## Discussion

**Chromosomal rearrangements are a source of standing variation**

We used paired-end Illumina sequence reads to identify chromosomal rearrangements in 14 sample strains derived from natural populations of *D. yakuba*. We identified genes at these locations they might affect. We identified 2368 rearrangement events within these lines of *D. yakuba*, indicating there is a substantial standing variation segregating in populations that may provide genetic material for adaptation.

Standing variation is expected to play a considerable role in evolutionary change and adaptive evolution (36). This variation provides the genetic diversity for a population to quickly adapt to new niches. We further provide evidence that there is significant variation in the presence and locations of rearrangements affecting the standing variation within populations.

Also, it appears that the genetic variation from rearrangements are dynamic complex. Some sites appear to have multiple rearrangement events and copy number changes are observed at some rearrangement breakpoints (Supplementary results). Further sequencing with long read technology would help advance the understanding of complex locations that are subject to multiple structural changes (35).

The conservative nature of our study offers a lower bound on the number of rearrangements that are in the genome. We required that rearrangements be supported with at least 4 abnormally mapping read pairs. There may be other mutations with lesser support that did not meet these thresholds. At least one case of a new gene being formed that did not meet the standards of our conservative approach, despite strong evidence in high coverage RNASeq data (S2 Figure). Hence, the full span of real biological variation is likely to be far richer than the limited portrait described here. Taken together this suggest that new gene formation and regulatory changes are an underestimated source of variation in natural populations.

**Chromosomal rearrangements are a source of new transcripts**

Previous theory has struggled to explain the ways that *de novo* genes might derive new open reading frames. The canonical progression of new gene formation suggests that many new genes appear as non-coding RNAs due to spontaneous gain of promoters to facilitate transcription (37). New transcripts would need to acquire translation signals to become fully formed new protein coding genes (9, 37, 38). Alternative explanations have suggested that pre-existing ORFs in the genome may be primed for translation even prior to transcription (6, 8). This mechanism raises the question of how translation signals might be recruited prior to transcription.

Here, we present evidence of *de novo* exons due to chromosomal rearrangements carrying promoters and translation start signals to new locations. New genes that result from such processes offer a clear genetic mechanism to explain new transcription. They also explain how translation signals can be acquired during *de novo* gene creation, changing expression and protein structure of new genes without multiple intermediary steps. The immediate progression to fully fledged coding sequences can explain how new genes form and how they might produce coding sequences without the need for secondary or tertiary mutations. With fewer mutational steps these genes are certain to form new proteins so long as translation start signals are captured. Hence, these mutations can explain the formation of new peptides without the possibility of loss through pseudogenization or deletion during protogene stages. We have identified 51 possible instances of the creation of *de novo* exons created from chromosomal rearrangements. Studies of tandem duplications uncovered over 100 combined new genes and 66 duplicated genes, suggesting that tandem duplications may affect gene novelty more than rearrangements in *D. yakuba* (21).

Genetic principles of rearrangements and new gene formation are likely to extend beyond the *Drosophila* model. At least one case of a new exon formation through rearrangement has been documented in humans where gene fragments drive expression on previously untranscribed regions (22). Hence, understanding of these genetic changes in model organisms is likely to offer important information that can be used for future studies on humans. Chromosomal rearrangements in humans are associated with cancers and infertility (39-42), and associated changes in gene copy number or chimera formation can influence risk of disease or evolutionary potential (43). Additionally, population diversity for new genetic content is essential to explain phenotypic variation within species in nature. Regulatory effects of gene relocation, new protein formation through chimeric genes, and *de novo* exon formation contribute to genetic changes across organisms. These genetic modifications, including new gene formation serve as a substrate of genetic novelty that is likely to be important for adaptation to new environments. As environments fluctuate, emerging new genetic material may become essential to facilitate phenotypic change. Surveys of standing variation in genome structure and gene content will therefore lead to better understanding of natural variation, adaptation, and disease.

**Chromosomal rearrangements are commonly associated with transposable elements**

Chromosomal rearrangements can be the result of multiple mechanisms including ectopic recombination, ectopic DNA repair or gene conversion, template switching during DNA synthesis, and transposable element movement. Transposable elements are a major mechanism of the rearrangements identified. We find 38% (909/2368) rearrangements have at least one TE within 1kb. Less than 5% of the major chromosome arms within *D. melanogaster* are transposable elements (44). Transposable elements have been hypothesized as a major players and catalysts in gene regulation networks (45). They often contribute sequence homology that can facilitate ectopic recombination. Yet, only 23.5% (12/51) rearrangements that may have formed a *de novo* exon are associated with TEs. This suggests that another mechanism such as gene conversion or ectopic recombination is responsible for the new genes formed. However, TEs and rearrangements could influence gene expression without changing the transcript. We found 134 genes that have significant differential expression from the reference within 1kb of identified rearrangements. Of these 134 rearrangements that are associated with genes, 74 (55%) are also associated with TEs. This suggests that TEs could be catalysts for the changes in gene expression in the genes that have altered expression in association with rearrangements.

**Genomic Distribution**

Sex chromosomes are subject to rapid rearrangement due to high repetitive content and selection to relocate gene content to autosomes. Consistent with these patterns, we observe an excess of rearrangements associated with the X chromosome in *D. yakuba*. We observe significantly more rearrangement sites on the X chromosome compared to the autosomes. This is consistent with previous findings that show the X chromosome has more structural variants in *Drosophila* (21, 46). In *D. melanogaster* the X chromosome has more repetitive content (47-49), unique gene density (50), and smaller populations size (49, 51). The X chromosome has lower levels of background selection, and contains an excess of sex specific genes (52, 53) compared to autosomes. Among rearrangements creating new transcripts we do not find an overrepresentation of breakpoints associated with the X chromosome, in contradiction with the "out-of-the-X" hypothesis of new gene formation (7, 26). Power may be limited to detect these effects with small numbers of new genes. Still, it is clear that X chromosome dynamics are unique, making it a prime resource to investigate the role of rearrangements in genome evolution

In addition to the excess of mutations on the X chromosome, we identify 4 'hotspots' of recurrent, independent mutation. Here, structural variants reshape variation at a single locus, with multiple low frequency variants segregating at the same region. The fact that single regions are mutated independently with unique breakpoints suggests either hypermutability or dynamics of selection on independent mutations similar to proposed 'soft sweeps'. A similar set of 'hotspots' has previously been noted for TE insertions at the locus of *klarsicht* in *Drosophila* (54) and in the evolution of pesticide resistance (55-57). Whether this locus represents a region subject to strong selection or is rather exceptionally labile remains to be determined. We observe mutations that rearrange sequences within chromosome arms rather than across independent chromosomes are proportional to the amount of DNA housed within the same chromosome arm. These results contrast with gene conversion data in mammals, showing that within-chromosome rearrangements are favored over cross-chromosome recombination during gene conversion (58).

## Methods

### Fly lines and genome sequencing

We used fastq sequences from previously published genomes (PRJNA215876, also available at https://drive.google.com/drive/u/0/folders/0Bxy-54SBqeekakFpeFBib3BXcVE) of 7 isofemale *Drosophila yakuba* lines from Nairobi, Kenya and 7 isofemale lines from Nguti, Cameroon (collected by P. Andolfatto 2002) (21). The reference strain is UCSD stock center 14021-0261.01, and the genome sequence is previously described in *Drosophila* Twelve Genomes Consortium (2007). Genome sequencing for the 14 isofemale lines are previously described in ref. 13. Briefly, the wild-caught strains and the *D. yakuba* reference stock were sequenced with three lanes of paired-end sequencing at the UC Irvine Genomics High Throughput Facility (http://dmaf.biochem.uci.edu).

### Sequence alignment and the identification of chromosomal rearrangements

We mapped paired-end genomic reads to the reference genome of *D. yakuba* r1.5 (20) and the *Wolbachia* endoparasite sequence (NC_002978.6) using bwa v/0.7.12 (59) using permissive parameters to allow mapping in the face of high heterozygosity in *Drosophila* (bwa aln -l 16500 -n 0.01 -o 2). The resulting paired-ends were resolved using “sampe” module of bwa to produce bam files. Each bam file was then sorted using samtools sort v/1.6 (59). These Illumina paired-end sequences were made with PCR amplified libraries (21). PCR duplicates can give false confidence in rearrangements through amplification of ligation products that do not represent independent DNA molecules. We used samtools rmdup to remove PCR duplicates. To identify genome structure changes, we used paired-end reads that were at least 1Mb away from each other or located on separate chromosomes (Figure 1). These abnormally mapped paired-end reads indicate possible rearrangements within or between chromosomes. Between 1 Mb and 100kb there may be some rearrangements but there are also inversions, moderately sized duplications (some with secondary deletions). A 1Mb threshold may exclude some variation, but allows greater clarity with respect to mutations and mechanisms that might generate mutations. We selected this stringent cut off to reduce the possibility of inversions being identified rather than translocations. To be considered as a possible rearrangement, a minimum of 4 independent reads must show the same paired-end read pattern (Figure 1). To be clustered together, sets of paired-end reads must be mapped within a distance smaller than the insert size of the library (325 bp) to each other on both rearrangement points. Only rearrangements involving major chromosome arms were considered. All heterochromatic or unplaced chromosomes were excluded.

Sequencing coverage is a major factor in false negative rates (21) (S8 Figure). When 4 supporting read-pairs are required to call mutations, we observe a strong correlation between depth and number of rearrangements ($R^2$=0.8223, $P$<4.8x$10^{-6}$) (S4 Figure). When only 3 supporting read-pairs are required to call mutations, there is no correlation between depth and rearrangement calls ($R^2$=0.009632, $P$>0.3) (S9 Figure). Four lines showed an unexpectedly large number of rearrangements when only 3 supporting read-pairs are used (S9 Figure). These sequence data were collected in early Illumina preparations before kit-based sequencing prep was available. Ligation of multiple inserts with high DNA concentration is likely to have produced this pattern. When 4 supporting read-pairs are required, the number of mutation calls fit into expected relative numbers between the lines negating the effects of errant insert ligation.

### Estimating false negatives and false positives

The number of structure calls is strongly correlated with depth of coverage of each line (S4 Figure). We estimated the number of reads of each line that would be expected at 93.7X coverage using a linear regression model between number read calls and depth of sequencing. In low coverage data, paired-end read may underestimate rearrangement numbers by as much as 50%. All flies sequenced were female. Hence, there should not be significant biases against identification of rearrangements involving the X chromosome compared to the autosomes. The lack of coverage in highly repetitive heterochromatic regions will limit ability to identify rearrangements at those loci. However, our goal was to find rearrangements that change gene structure or expression, while heterochromatic regions are generally less gene dense. Requiring 4 supporting read-pairs may lead to many false negatives in low coverage data. To identify specific cases of false negatives, we surveyed each confirmed rearrangement in each line that did not have a positive call that rearrangement. If these other sample lines had 1-3 reads supporting a rearrangement we considered it a false negative in that sample strain. This is expected to identify many of the false negatives, but there may be false negatives that may not have had a single abnormally paired read supporting it which we failed to identify.

False positive rates were determined by using previously published long read PacBio sequences (21). PacBio sequencing was done for 4 lines NY73, NY66, CY17C, and CY21B3. This sequencing experiment was done in the early stages of long read sequencing and thus coverage depth for each line is between 5X and 10X. We matched PacBio sequence reads to the *D. yakuba* reference using a BLASTn with the repetitive DNA filter turned off and an *E*-value cutoff of $10^{-10}$. If a single molecule read matched in a BLASTn within 2kb of both sides of the genomic rearrangement call it was considered confirmed. The number of rearrangements that were not confirmed divided the number of the total rearrangements for that line provides us with an estimate of the false positive rate.

**Polarization of the ancestral state**

All the rearrangements identified are polymorphic in populations and are expected to be relatively new changes. However, each rearrangement was determined relative to the reference strain. Therefore, it is possible that the rearrangement identified could represent a new rearrangement or the ancestral state that has been rearranged in the reference strain. To polarize the rearrangements, we acquired sequences 1kb upstream and 1kb downstream of each rearrangement site. These sequences were then matched to the *D. erecta* reference genome using a blastn (20).

If the two sides of a rearrangement aligned within 2kb of each other on the same chromosome in *D. erecta*, it was determined that the rearrangement call is the ancestral allele and the reference has the derived allele. Rearrangements that are shared across species will accumulate nucleotide differences. Therefore, hits must have a minimum of 85% nucleotide identity and must span at least a segment of rearrangement call breakpoints as defined by abnormally mapping Illumina reads. Rearrangements are commonly associated with transposable elements and repetitive element, so if the two sides of a rearrangement map close to each other in more than 10 locations the ancestral state could not be determined.

**Gene expression changes**

We used previously published RNA sequences (13, 14) to identify gene expression changes and new gene formation associated with genome structure changes. Briefly RNASeq samples were prepared from virgin flies collected within 2 hrs. of eclosion, then aged 2-5 days post eclosion before dissection. Available data includes ovaries and headless carcass for adult females, and testes plus accessory glands (abbreviated hereafter as testes) and headless carcass for adult males. Sequence data are available in the NCBI SRA under PRJNA269314 and PRJNA196536.

We aligned RNASeq fastq data to the *D. yakuba* reference genome using Tophat v.2.1.0 and Bowtie2 v.2.2.9 (60). We utilized Tophat-fusion search algorithm (61) to identify transcripts that represent fusion gene products either between chromosomes or rearrangements within chromosomes. To confirm fusion events, RNASeq fastq data were assembled reference-free into a transcriptome using Trinity v.2.4.0 (29). Each transcriptome was then matched to the *D. yakuba* reference using a BLASTn with the repetitive DNA filter turned off and an *E*-value cutoff of $10^{-10}$. All genomic mutations are identified as differences between sample and reference strains. Hence, the RNAseq coverage in the reference serves as a 'control' to help identify new genes formed at rearrangement breakpoints.

**Identifying fusion transcripts and gene expression changes**

Genomic rearrangement calls were matched to fusion calls from Tophat fusion (61) for testes, ovaries, male carcass, and female carcass. If the two sides of a supported rearrangement were within 1kb of the three Tophat fusion reads or read-pairs (Figure 2), the rearrangement was considered candidate *de novo* exons. Genes annotations in *D. yakuba* r1.5 within 1 kb of each location of the RNA supported genomic rearrangement calls were identified. Rearrangements where one side is located near a gene and the other side is not, were of particular interest for the creation of *de novo* exons.

Gene expression at each rearrangement was quantified using coverage depth divided by total mapped reads, analogous to FPKM correction. Each of the four tissues described above (testes, male carcass, ovaries, female carcass) were screened for sequence expression differences associated with the rearrangements. Regions that have unique expression patterns associated with rearrangement calls are considered new transcripts. When a rearrangement brings together a gene and a noncoding locus and there is new transcription in the noncoding region is indicative of new genes.

These new genes were further confirmed using the reference free transcript assembler, Trinity. Each transcript was compared to the *D. yakuba* references using BLASTn with the repetitive DNA filter turned off and an *E*-value cutoff of $10^{-10}$. Transcripts that matched to both ends of the rearrangement was considered confirmation.

We used previously published data (18) from the Cuffdiff program of the Cufflinks differential expression program (30) to search for regulatory changes in genes near chromosomal rearrangements. These data rely on previously published gene and transcript annotations from the same RNASeq data (62). We compared gene expression of each gene versus the reference strain. Genes that were within 1kb of a chromosome rearrangement call and had significant change from the reference strain were identified.

**Gene ontology**

Gene ontology was analyzed using DAVID GO analysis software (http://david.abcc.ncifcrf.gov) (63, 64). We surveyed for overrepresentation of genes within differing functional pathways. Functional groups with an enrichment score greater than 2 were reported. Functional genetic data for *D. yakuba* remains sparse. To determine functional categories represented, we identified *D. melanogaster* orthologs as classified in FlyBase and used these as input for gene ontology analysis.

**Differences between chromosomes**

We analyzed differences among chromosomes using an ANOVA and Tukeys HSD tests using random block design using line as the treatment blocks. To tabulate rearrangement sites among the chromosomes, each rearrangement that was within a singular chromosome arm counted as 2 sites on that chromosomal arm while rearrangements between chromosomes counted as 1 site on each of the chromosomes involved. Differences between the chromosome arms involving rearrangements within a chromosome arm and between chromosome rearrangements were identified individually using an ANOVA and Tukeys' HSD tests using the same random block design.

**Population genetics**

Estimates of $\theta_\pi$, $\theta_\omega$, and Tajima's D in 5kb windows for this of D. *yakuba* (https://github.com/ThorntonLab/DrosophilaPopGenData-Rogers2015) were previously described in ref 46. These estimates excluded sites with missing data, ambiguous sequence, or heterozygous sites. We report population genetic statistics for each window containing rearrangements and new genes in the data presented here.

Table 1: Number of chromosomal rearrangements found in the 14 lines. Our ability to identify rearrangements was associated with sequence coverage depth. Number of rearrangements for each line was then predicted at coverage depth of 93.7X coverage using a linear regression model.

| Line | Coverage depth | Between chromosomes | Within chromosomes | Total | Between chromosomes (Predicted at 93.7X) | Within chromosomes (Predicted at 93.7X) | Total (Predicted at 93.7X) |
|---|---|---|---|---|---|---|---|
| *NY73* | 27.6 | 129 | 57 | 186 | 346 | 129 | 475 |
| *NY66* | 26.5 | 111 | 47 | 158 | 332 | 120 | 452 |
| *NY62* | 44.8 | 177 | 74 | 251 | 337 | 127 | 464 |
| *NY48* | 34.5 | 156 | 59 | 215 | 351 | 124 | 475 |
| *NY56* | 12.1 | 70 | 26 | 96 | 338 | 115 | 453 |
| *NY81* | 23.9 | 119 | 47 | 166 | 348 | 123 | 471 |
| *NY85* | 61.1 | 214 | 82 | 296 | 321 | 118 | 439 |
| *CY22B* | 45.5 | 148 | 60 | 207 | 306 | 113 | 419 |
| *CY21B3* | 44.8 | 155 | 74 | 229 | 315 | 127 | 442 |
| *CY20A* | 93.7 | 321 | 102 | 423 | 320 | 102 | 422 |
| *CY28A4* | 58.3 | 221 | 105 | 326 | 337 | 143 | 480 |
| *CY04B* | 64.3 | 326 | 129 | 455 | 422 | 161 | 583 |
| *CY17C* | 43.2 | 171 | 73 | 244 | 337 | 128 | 465 |
| *CY08A* | 37.5 | 190 | 70 | 260 | 407 | 142 | 549 |

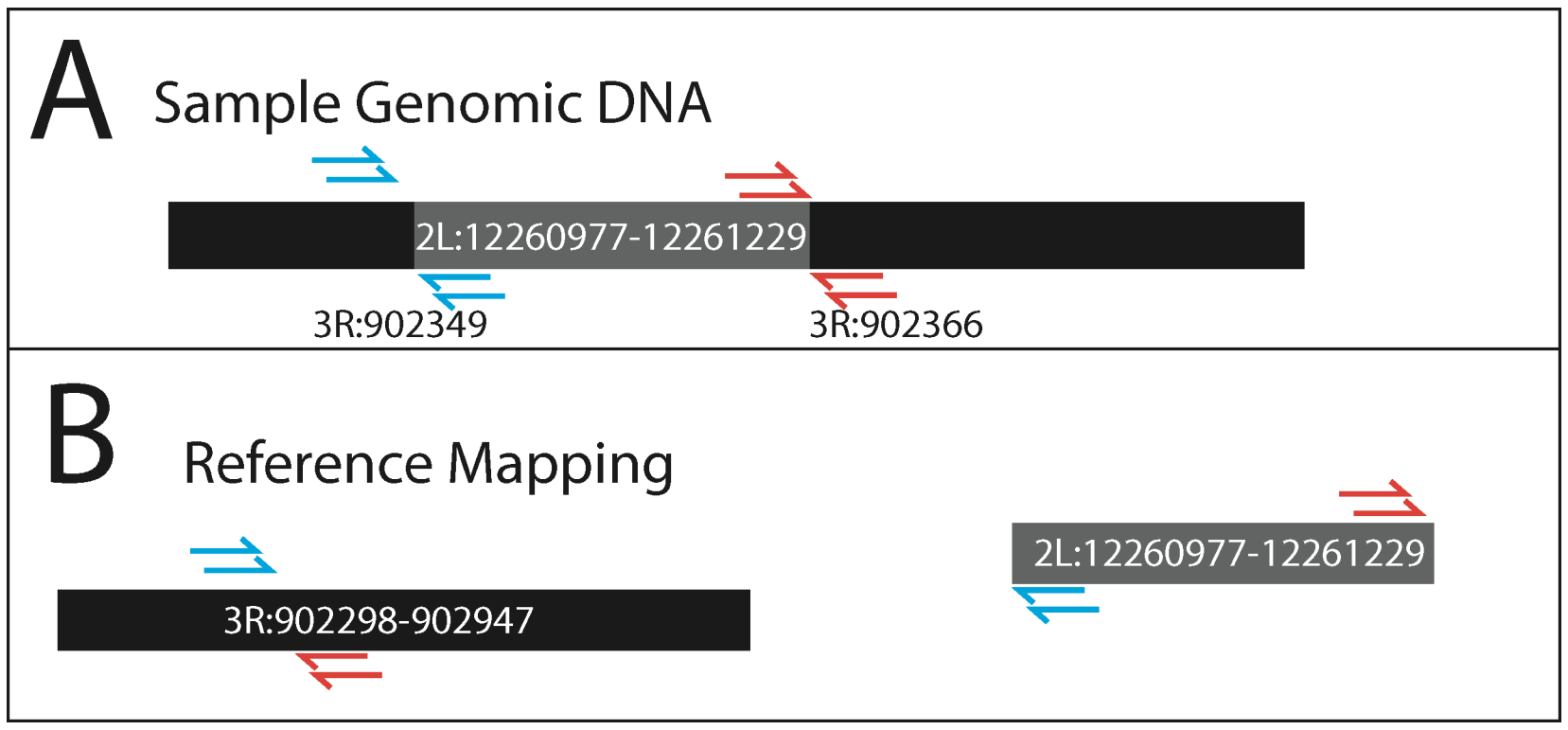

Figure 1: Example of paired end reads mapped abnormally to the reference genome. A) CY17C Chromosome sequenced with paired reads with 325 bp insert size. CY17C has an insertion of sequence from 2L into chromosome arm 3R. B) Each end was then aligned to the reference genome. Left reads that mapped around 3R:902100 had paired right reads mapped to regions near 2L:12261000 (red arrows). Additionally, right reads that mapped around 3R:902600 paired with left reads mapping around 2L:12261150 (blue arrows). This indicates that the region between 2L:12261000-12261150 has been inserted into 3R: 902000-902600 in line CY17C. Each rearrangement needs at least 4 abnormally mapping read pairs to be considered.

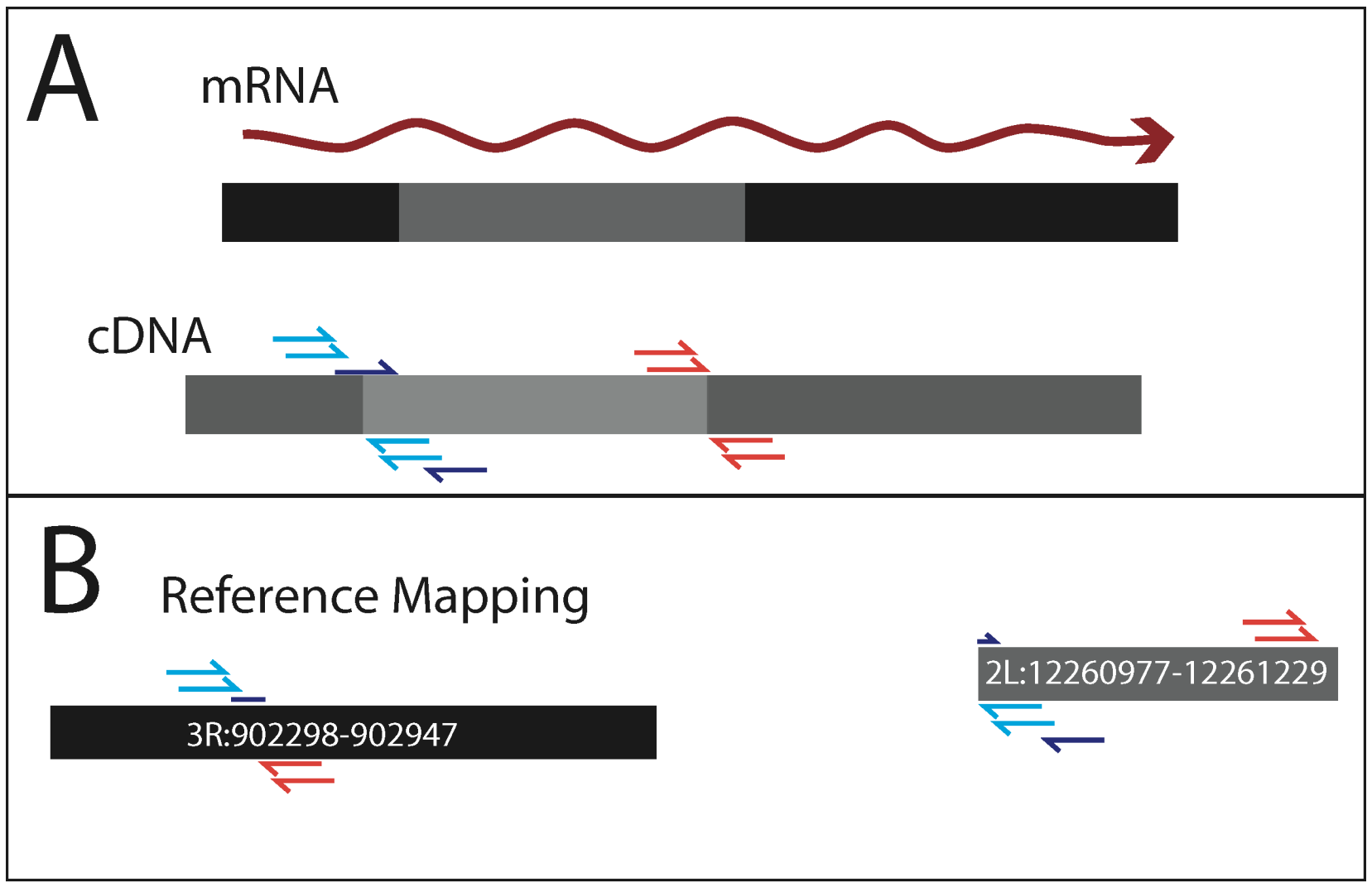

Figure 2: Example of paired end sequence reads mapping from RNASeq data. A) CY17C has an insertion of sequence from 2L into chromosome arm 3R. This insert placed a previously untranscribed region within a previously transcribed gene. Paired end reads were generated from cDNA. B) The paired end reads of this RNA transcript will map to separate chromosomes on the reference sequence, and split read mapping may be seen at the breakpoints. Three total misaligned RNASeq pairs and/or split reads is needed to be considered a formation of a new gene.

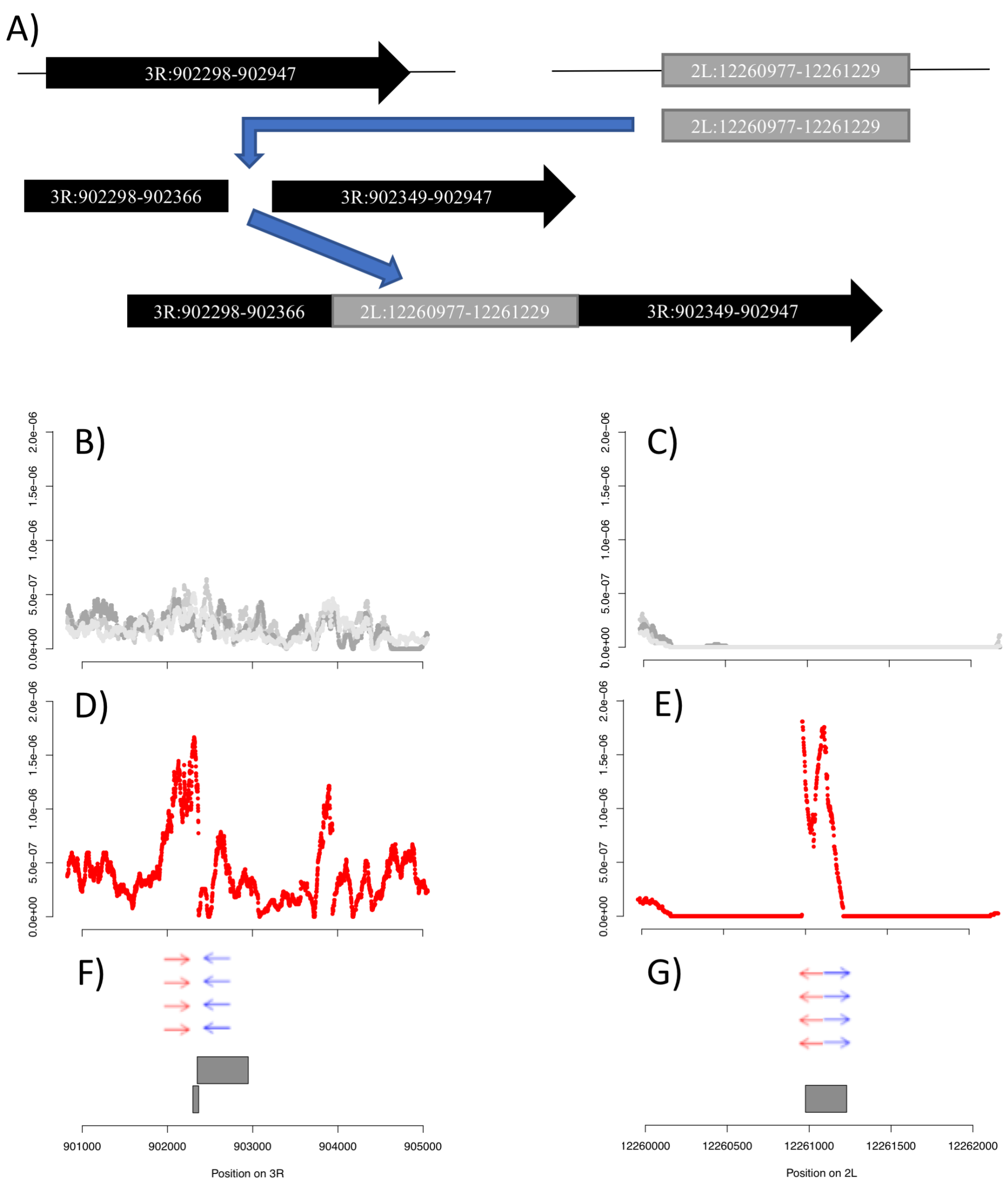

Figure 3: New gene formation through genome rearrangement on chromosome 3R and 2L. Observing sequence depth of the RNA we can infer relative expression and identify newly transcribed regions in lines that have rearrangement calls. Relative RNA Coverage depth was calculated from Tophat RNASeq alignments by dividing the read depth at each base by the total number of reads mapped. Two regions that have 2 genomic rearrangement calls and Tophat fusion calls supporting the formation of a *de novo* gene. A) Diagram showing the predicted

sequence movement based on the Trinity Transcript blast. An insertion of the sequence from 2L:12260976 in-between 902154 and 902563 has moved a segment of previously untranscribed DNA to a region with active transcription on 3R. RNA transcript assembled by Trinity confirms the observed coverage pattern in RNASeq data. The transcript starts near 3R:902000, the middle section mapped between 2L:12260976-12261178 and the final section then maps near 3R:902500. B) and C) The grey coverage lines are RNA sequence coverage from 3 reference RNASeq replicates which do not have this rearrangement. D) and E) RNA sequence coverage of line CY17C which has the rearrangement present. F) and G) CY17C has a two genomic rearrangement calls between 2L:12260976-12261229 matching with 3R:901825-902154 (red arrows) and 3R:902563-902607 (blue arrows). Grey boxes represent the Trinity transcript aligned to the reference genome.

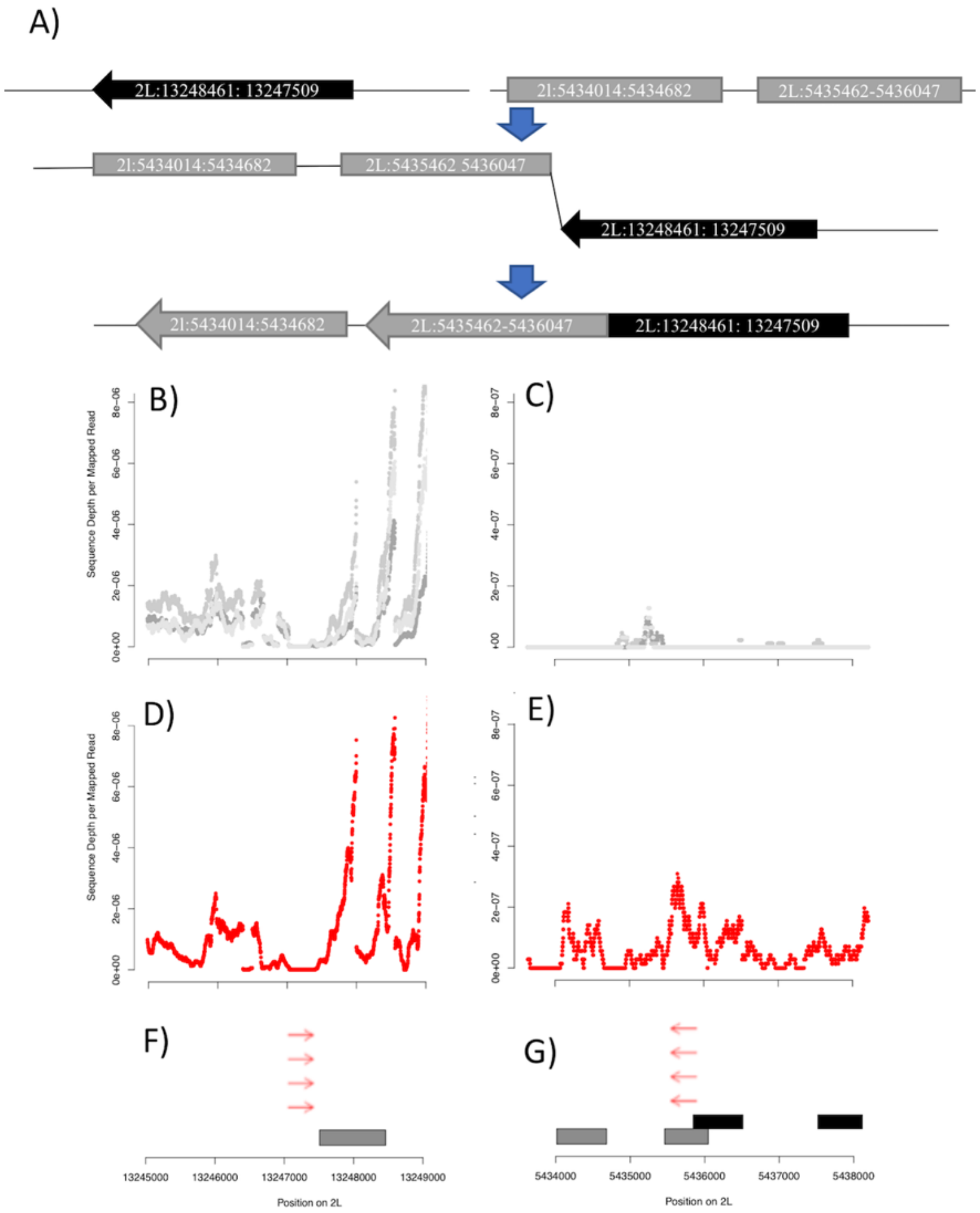

Figure 4: A) Diagram showing the predicted sequence movement based on the Trinity Transcript blast. A rearrangement joining the sequence from 2L:5435633-5436212 902154 and 2L:5435462-5436047 has moved a segment of previously untranscribed DNA to a region with active transcription on 2L. B) and C) The grey coverage lines are RNA sequence coverage from 3 reference RNASeq replicates which do not have this rearrangement. D) and E) CY28A4 has the rearrangement and increased transcription in region 2L:5435462-5436047. F) and G) CY28A4 has a rearrangement calls between 2L:13246986-13247746 matching with 2L:5435633-5436212 (red arrows) and Square boxes in represent the Trinity transcript aligned to the reference genome. The black boxes represent exons of a preexisting gene (1.g484.t1**)**

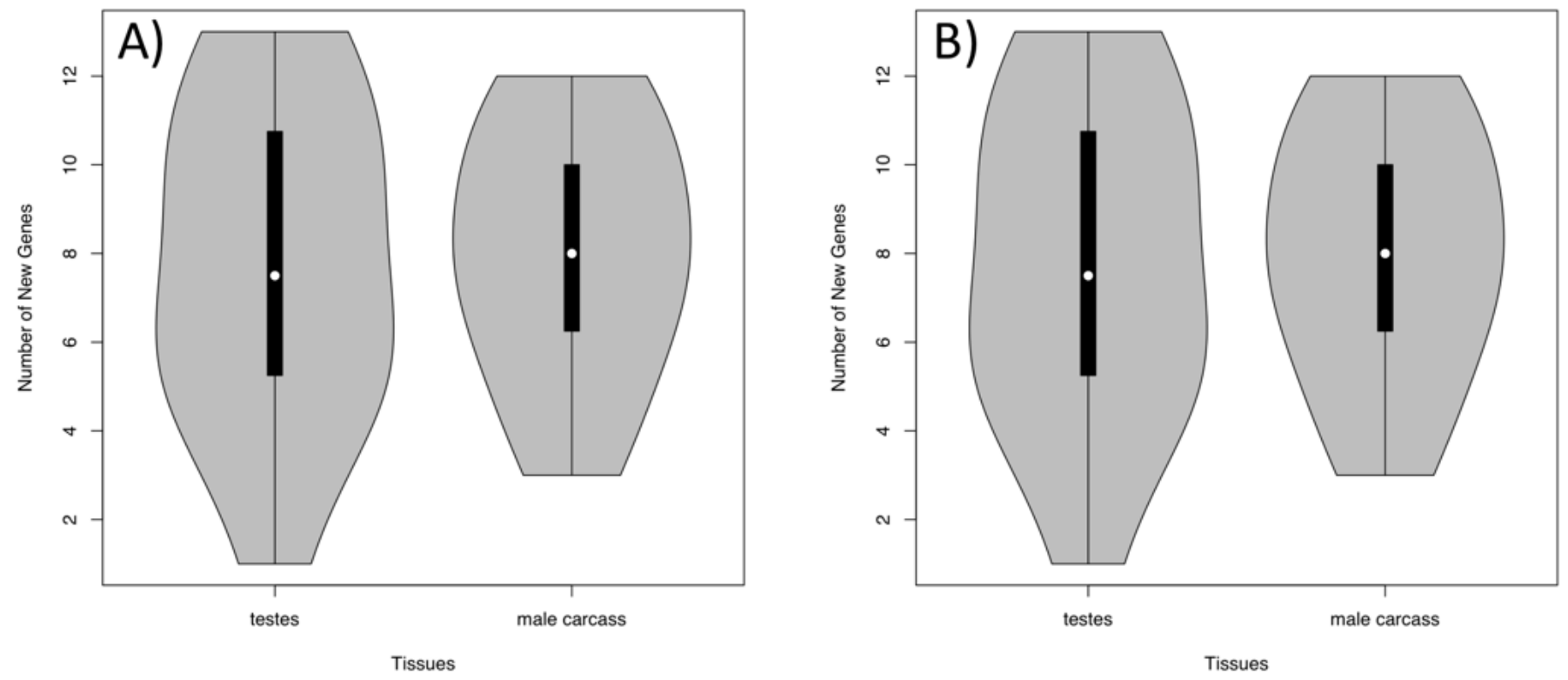

Figure 5: Distribution of new genes per strain identified in testes, male carcass, ovaries, and female carcass based on 14 inbred lines in males (A) and females (B). A total of 51 new genes were identified across all 14 strains in all tissues. We not see a difference in the number of new genes expressed between male gametic and somatic tissue (ANOVA, $F(1,13)=0.04$, $P>0.8$). While there is a significant difference between ovaries and female tissue (ANOVA, $F(1,13)=4.379$, $P<0.05$), the values are low for each line (including being 0 for multiple samples). This suggest that the male comparison is more indicative of the ratio between somatic and gametic tissue.

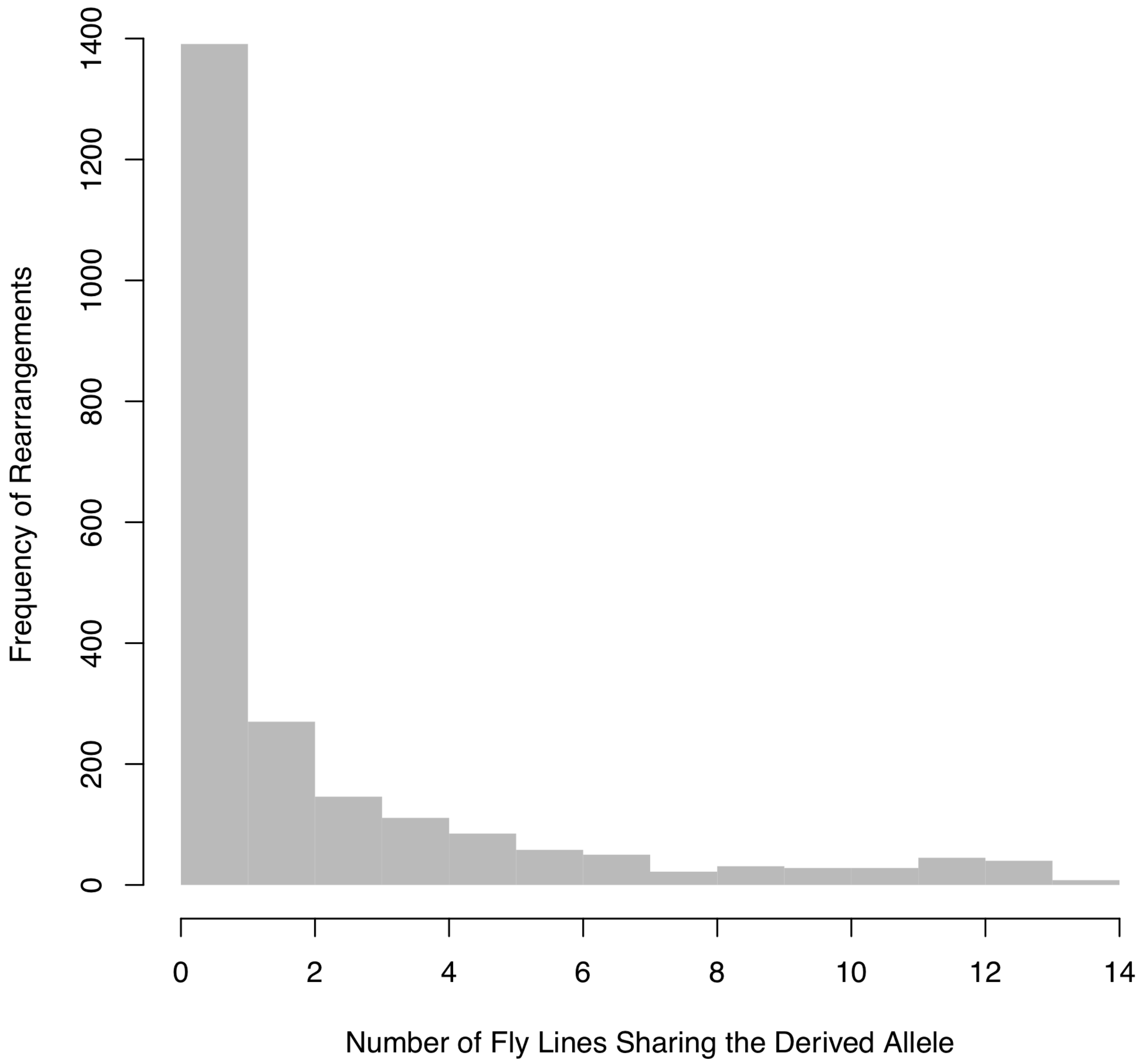

Figure 6: Site frequency spectrum of rearrangements found in the 14 lines. Most of the rearrangements are singletons. However, there is a slight increase in number of rearrangements found in at least 11 of the 14 lines.

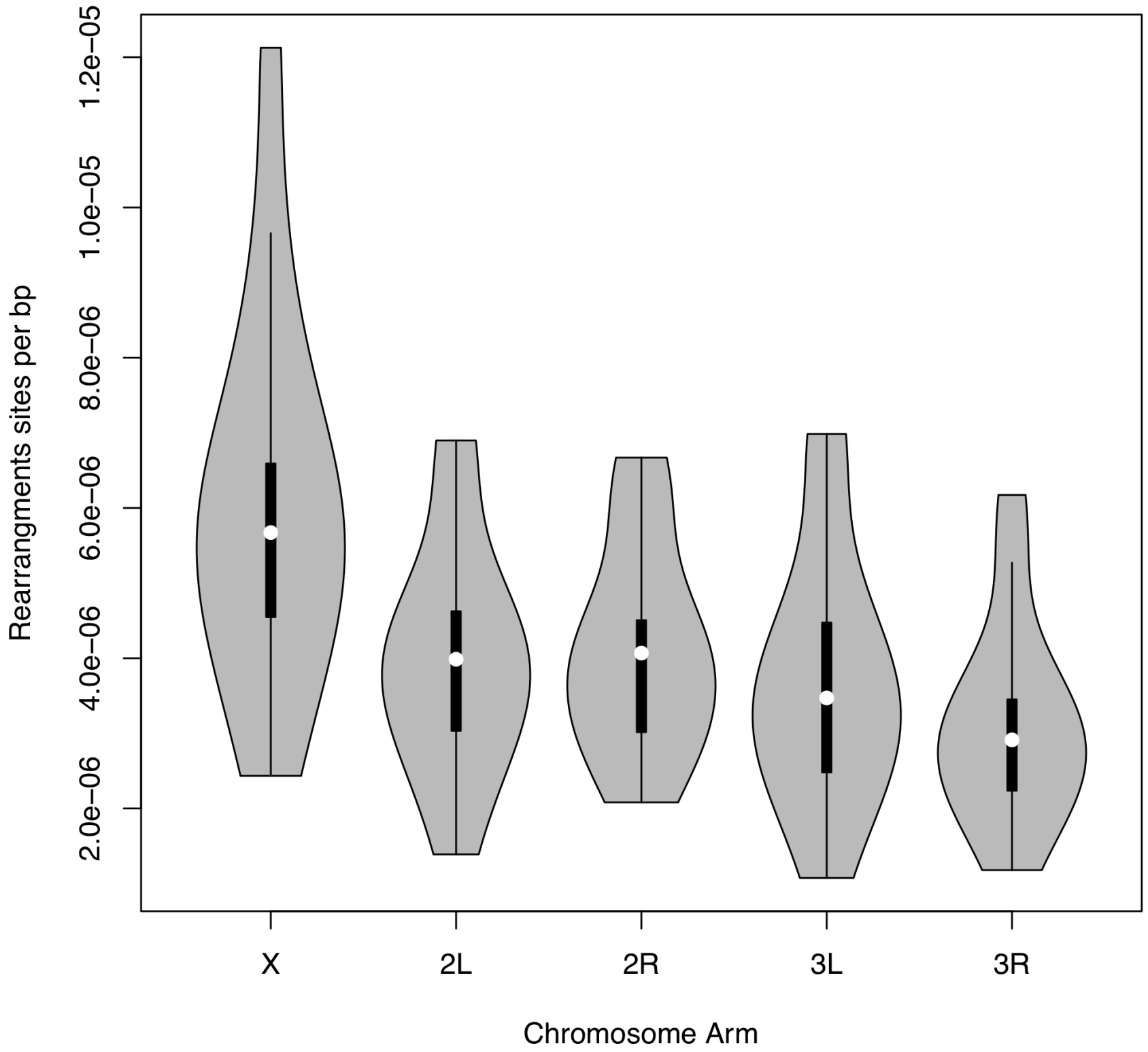

Figure 7: Number of rearrangement breakpoints per base pair on each chromosome arm for inbred lines of *D. yakuba*. Total number of rearrangement sites on each chomosome varied (ANOVA, $F(4,52)=43.42$, $P<10^{-15}$). This is mostly do to the the fact that the X chromosome has significantly more rearrangement breakpoints than the autosomes (Tukey HSD for each comparison involving the X, $P<10^{-6}$). Chromosome 3R had significantly fewer rearrangements than the X, 2L and 2R (Tukey HSD, $P < 0.05$).

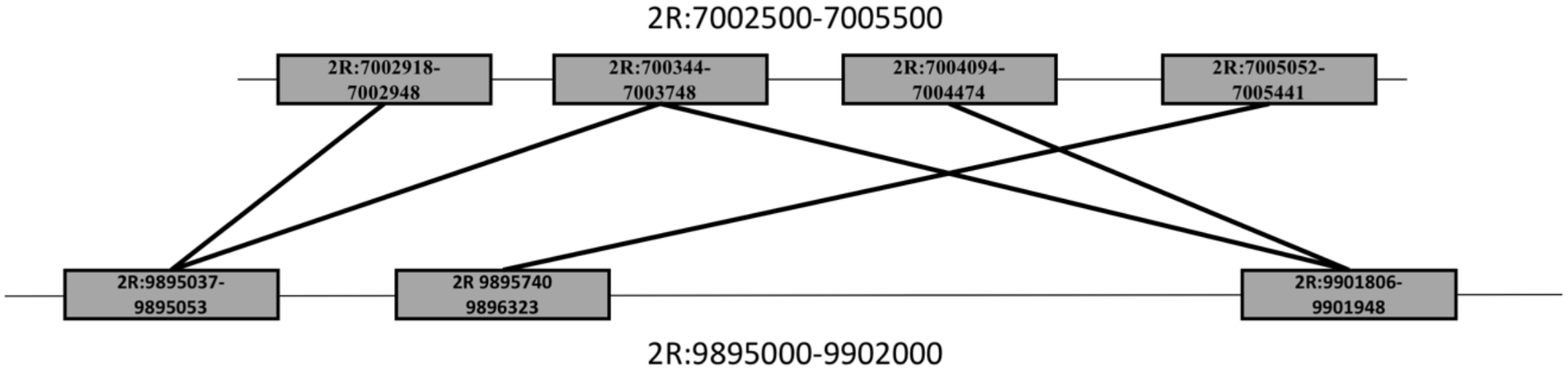

Figure 8: Many rearrangements lie in the same region making it hard to fully elucidate the nature of a particular rearrangement. For instance, in line CY21B3 has 5 rearrangement calls (represented by the connecting lines of the two large sections of chromosome 2R) associated with two regions 2R:7002500-7005500 and 2R:9895000-9902000. 4 separate small regions that are separated by at least 1 sequencing insert size (325 bp) within 2R:7002500-7005500 have reads that pair with 3 separate small regions between 2R:9895000-990200. All the lines always show at least one of these rearrangements but generally each line has 2-3 separate rearrangement calls between regions 2R:7002500-7005500 and 2R:9895000-9902000. The 9.9Mb breakpoint lies close to the known inversion breakpoint on 2R where recombination is suppressed.

**S1 Text**

**Supplementary Information**

**Types of rearrangements**

We classified the rearrangement identified as putative translocations. These genetic changes may encompass any mutation that moves DNA from one location to another. They will include ectopic gene conversion, ectopic recombination, TE movement, repair of staggered breaks, template switching during replication, and putatively retrogenes formation. We classified the rearrangements that were associated with transposable elements, duplications, and inversions

**Duplications associated with rearrangements**

Transposable element proliferation and duplicative rearrangements can increase genome content by creating extra copies of DNA sequences. To determine the propensity of these rearrangements to create duplicates, we calculated the average depth of coverage of each rearrangement plus 100 bp upstream and downstream and 500 bp upstream and downstream of each rearrangement. We classified possible duplications associated with rearrangements by identifying the rearrangement sites that have average sequence coverage depth that is twice the average coverage depth of that line. Duplications are associated with 5-25% of the rearrangement sites within each line (S6 Table). Line NY56 has an elevated number of possible duplications, however this is most likely associated with the low average coverage (12.1X) skewing the results. Overall, we observe that the majority of the rearrangement sites are not associated with duplications (S6 Table).

More rearrangement sites appear to be duplicated in 100 bps flanking the rearrangements than 500 bps regions flanking the rearrangement sites and as the sequence depth decreases moving away from the rearrangement. The observed coverage differences suggest increased coverage more immediate to the rearrangement suggesting more small duplications around the rearrangement sites than large ones. These results are consistent with other studies suggesting *D. yakuba* has many small tandem duplications (21)**.** It may also be consistent with target site duplications at breakpoints as has previously been observed for inversions (34).

**Association with inversions**

Previously, 29 inversions were identified in *D. yakuba* (32). We identified rearrangement calls from our dataset that were within 500KB of the estimated breakpoint region of each inversion. Overall, we found 5.3% (36/671) of the rearrangements within a singular chromosome matched to previously identified inversions. Additionally, our rearrangement calls only associated with 55% (16/29) of the previously identified inversions. The methods described here were not intended to identify inversions, which often have complex rearrangements and repeats at their breakpoints. Furthermore, it is possible that some inversions are not present among the strains sequenced here.

Polymorphic inversions within *D. yakuba* would appear as possible rearrangements within our study. In the case of an inversion, the misaligned read pairs would mark inversion break with read pairs would be mapped in the same direction. We do see evidence that within each line roughly 40-60% (mean 51%) (S7 Table) of the rearrangements on the same chromosome do have read pairs that are mapped in the same direction. However, this could be an indication of a translocation when the insert was inserted in reverse rather than a traditional inversion with two breakpoints encompassing all intervening sequence. We note that these numbers are consistent with a 50/50 split of strand orientation for translocations. Rearrangement between chromosomes cannot be inversions. These show that within each line a similar proportion of 40-55% (mean 47%) of rearrangements having reads in the same direction. If we identify many inversions, then we would expect a higher proportion of the same read pairs in rearrangement within rearrangements than between chromosome structure calls. This combined with that overall roughly 5% of our rearrangements match previously identified inversions suggests that the majority of the rearrangements that have been identified were most likely not inversions. Previous analysis of polymorphism did not note signals consistent with these mutations representing inversions on chromosome arms but did identify atypical SNP patterns in regions matching known inversions on 2R (32, 34).

Previous work on these sequence data estimated that sequencing under 30X coverage would significantly reduce the effectiveness of identifying tandem duplications (21). Our results suggest that this also applies with chromosomal rearrangements. The number of chromosomal rearrangements identified is significantly correlated with sequence depth ($R^2$=0.8231, P<$4.7x10^{-6}$) (S4 Figure). We used a linear regression to correct the total rearrangements identified in each line based on coverage (Table 1). After corrections for coverage, we would expect between 419 and 583 rearrangements per strain at 94X coverage (Table 1).

Previously published PacBio sequencing was available for four of the lines: NY73, NY66, CY21B3, and CY17C (21). PacBio sequence reads that match using BLASTn within 2kb of both rearrangement sites were considered confirmation of the rearrangement. Between 80%-97% of the rearrangements per strain identified using paired-end Illumina sequence reads could be confirmed in the PacBio long molecule sequencing. Confirmation rates vary according to sequence coverage depth of the PacBio sequencing ($R^2$=0.9883, $P$<0.004) (S4 Figure). The confirmation rate of each line is not correlated with the sequence depth of the Illumina sequencing, implying that false positive rate from paired-end read mapping does not depend on coverage. To further analyze false positive, we analyzed rearrangements that are associated with transposable elements at both breakpoints. Transposable elements have the best likelihood to create false positives by misalignments. Overall in these four lines, PacBio reads confirmed at 88% of such rearrangements in each line. The high number of confirmed rearrangements suggests that the number of false positives identified by Illumina sequencing may be under 5% of the total number of rearrangements identified, consistent with previous work (21, 22).

**Quality Control for new gene formation**

To further evaluate differences between paired end and single end sequencing, we down sampled RNAseq data for males to determine performance using single end reads. Our rearrangement calls using single end reads are reduced from 43 to 21 in testes and from 42 to 16 rearrangements in male carcasses, suggesting much greater power in paired-end read data. New gene formation per line varies from 3 new transcripts (NY48) to 16 (NY62, NY66). No *de novo* transcripts exist that were expressed in ovaries or female somatic tissue that were not also expressed in either testes or male somatic tissue. It is likely that lack of female-specific genes is due to lower power to identify fusion transcripts, though previous work has noted *de novo* genes in *Drosophila* are male-biased (7, 8). We identified putative false negatives as rearrangements in lines with less than 4 support reads that match a rearrangement call in another line. This analysis reveals another 6 constructs, for a maximum of 58 new genes. Additionally, we identified one locus that fails to meet the threshold of 4 independent Illumina sequences, but appears to be a new gene with strong evidence in high coverage RNASeq data (S2 Figure). This new locus shows strong upregulation matching the inferred gene structure from a putative rearrangement supported by 3 read-pairs in genomic data.

Of the 39 transcripts that were confirmed with Trinity, possible start codons were located before the breakpoint in 34 (87%) of the transcripts. Of the 13 rearrangements that were not confirmed with Trinity, 6 of them fell within a previously annotated gene. In all 6 cases the start codon appears to be before the rearrangement breakpoint. Of the other 7 rearrangements 4 appear to have break points within 300 base pairs of previously annotated genes. However, we cannot pinpoint the exact breakpoint of the rearrangements to single bp resolution. This could indicate that regulatory elements have moved but not exons. The final 3 rearrangements, neither break was within 1000 base pairs of a known gene. This could indicate that this rearrangement created a *de novo* gene without the incorporations of existing exons or an incomplete annotation within the *D. yakuba* genome.

**Genomic Distribution of Rearrangements**

We wondered whether there might be 'hotspots' for rearrangements across the genome. Four hotspots were identified: one on the X, one on 2R, and two on 3R (S7 Figure). One was classified as having over 30 rearrangement breakpoints across 14 lines within a 5kb window (e.g. Fig 8). Interestingly, these rearrangement hotspots were not specific to a particular line. Each hotspot had structure calls associating it with a range of locations and each line has less than 10 rearrangements associated with each region. In aggregate, the population genomic data suggests repeated, independent mutations affecting the same regions in different strains. Most of the rearrangements associated with hotspots are singletons in the data set. The hotspot on 2R close to position 10Mb lies adjacent to a known inversion breakpoint which has been shown to suppress recombination (34).

Multiple rearrangements in the same region suggested action by transposable elements that have independently moved in the lines to unique locations. None of the four rearrangement hotspots were associated with TE calls in a previously reported dataset (33). However a BLAST comparison of sequences to Repbase (65) showed high matching with a TE family including *CMC-Transib*, *Mariner*, and *Jockey* families. Several factors may explain these previously unreported TE calls. Mapping criteria that allow for a greater number of mismatches to capture heterozygosity may allow for additional read mapping than previous strict matches. New releases of bwa may process alignments differently. Also, previous TE calls required breakpoint assembly of the TEs to classify the locus as a transposable element (33), while analysis here does not. The four hotspots all showed similar trends suggesting that highly complex cases of TE movement may have breakpoints that are difficult to assemble in short read data.

**Population Genetics of Rearrangements**

Paired-end read mapping identifies rearrangements based on differences compared with the reference. It does not identify ancestral states. To determine whether mutations were derived or ancestral, we used BLASTn to compare sequences 1kb downstream and 1kb upstream of each rearrangement site to *D. erecta.* If the two regions matched a single region within 2kb of one another in *D. erecta* then the derived allele is present in the reference while the sample strains contain the ancestral state. We identified 112 rearrangements or 4.7% of the rearrangements that represent new mutations in the *D. yakuba* reference rather than in sample strains. A total of 54 out of 2368 of the rearrangements had pairs that matched over 10 different sites making it too difficult to confidently identify the ancestral and derived allele in the face of repetitive DNA. These were excluded from the site frequency spectrum (SFS).

Of our 2368 rearrangements, we identified 167 (7.1%) have at least one of the rearrangement sites associated with a Tajima's *D* in the bottom 5% (-2.27) of Tajima's *D* measure throughout the major chromosome arms. However only 170/4621 (3.7%) sites have a Tajima's *D* of less than -2.27. We used a bootstrap approach to determine whether sweep-like signatures were overrepresented at these rearrangements. We randomly sampled windows and recorded Tajima's *D* 4621 times along the major chromosome arms with 10,000 replicates. All replicates had at least as many random windows with Tajima's *D* less than -2.27, suggesting rearrangements are underrepresented in regions with signatures of strong, recent selection. This suggest that these rearrangements are most likely to be deleterious or neutral. Our ability to detect selection on our samples might be diminished if hotspots, which vary across lines, bias results to regions with high amounts of segregating variation. It is also possible that these sites could be subject to soft sweeps, which are not easily detected by reduced diversity or site frequency spectra skewed toward rare alleles (66).

While certain rearrangements identified with very low Tajima's *D* are new singleton rearrangements, our analysis could be identifying artifacts of selective sweeps before the appearance of the rearrangement. Derived rearrangements found at high frequencies are may be candidates for selectively favored variation in the population. We identified 125 derived rearrangements found in at least at least 75% (11/14) of the lines. These rearrangements are within 1 kb of 78 genes, 52 of which have orthologs in *D. melanogaster*. These genes have functions in many functional groups including Rho GTPase activity and imaginal disc and wing vein morphogenesis. Of 125 rearrangements 10 (8%) rearrangements had at least one side of the rearrangement has a Tajima's *D* in the lowest 5% (Tajima's $D < -2.27$). However, the overall number of rearrangement sites found in at least 75% of the lines are not overrepresented (11/241, 4.6% have Tajima's $D < -2.27$). Random sampling of 241 windows along all the chromosomes with 10,000 bootstrap replicates reveals that our sample is well in the expected range (1-28) with 30-40% of the trials match or have fewer sites within the bottom 5% of Tajima's *D*.

**Gene ontology**

To characterize the functional categories that might be affected by these mutations, we explored gene ontology (GO) categories of genes associated with the rearrangements. We used DAVID GO analysis software (63, 64) to investigate the genes being altered by chromosomal rearrangements. We identified 1202 prospective genes within 1kb of a rearrangement call. We identified 733 genes that have a *D. melanogaster* ortholog and are within 1 kb of a rearrangement site. Overrepresented functional categories include alternative splicing, transmembrane proteins, protein phosphorylation, and glycoproteins (S8 Table).

We identified 52 genes within 1 kb of the estimated breakpoints of structure calls that are supported by RNASeq Tophat fusion calls. These loci are candidates for new gene formation. A total of 37 of these genes have orthologs in *D. melanogaster*. These genes however showed no overrepresentation with respect to gene ontology or functions.

**S1 Table:** New gene formation in strains of *D. yakuba*. Raw totals of chromosomal rearrangements on each chromosome arm that are supported by genomic structure calls and Tophat fusion calls for testes, male carcass, ovaries, and female carcass.

| line | X | | 2L | | 2R | | 3L | | 3R | | 4 | | Total |
|---|---|---|---|---|---|---|---|---|---|---|---|---|---|
| | **within** | **between** | **within** | **between** | **within** | **between** | **within** | **between** | **within** | **between** | **within** | **between** | |
| NY73 testes | 0 | 1 | 1 | 2 | 2 | 0 | 1 | 0 | 0 | 1 | 0 | 0 | 6 |
| NY73 mal car | 0 | 0 | 1 | 1 | 2 | 0 | 1 | 0 | 0 | 1 | 0 | 0 | 5 |
| NY73 ovary | 0 | 1 | 0 | 0 | 0 | 0 | 1 | 0 | 0 | 1 | 0 | 0 | 2 |
| NY73 fem car | 0 | 0 | 0 | 1 | 0 | 0 | 0 | 0 | 0 | 1 | 0 | 0 | 1 |
| NY66 testes | 0 | 3 | 3 | 1 | 0 | 0 | 1 | 5 | 3 | 1 | 0 | 2 | 13 |
| NY66 mal car | 0 | 1 | 1 | 2 | 1 | 0 | 1 | 2 | 0 | 0 | 0 | 3 | 7 |
| NY66 ovary | 0 | 1 | 0 | 0 | 0 | 0 | 1 | 2 | 0 | 1 | 0 | 2 | 4 |
| NY66 fem car | 0 | 0 | 0 | 0 | 0 | 0 | 1 | 0 | 0 | 0 | 0 | 0 | 1 |
| NY62 testes | 0 | 5 | 0 | 3 | 3 | 1 | 0 | 2 | 3 | 2 | 0 | 1 | 13 |
| NY62 mal car | 0 | 3 | 0 | 5 | 1 | 0 | 0 | 3 | 0 | 1 | 0 | 0 | 7 |
| NY62 ovary | 0 | 1 | 0 | 0 | 1 | 0 | 0 | 0 | 0 | 1 | 0 | 0 | 2 |
| NY62 fem car | 0 | 0 | 0 | 0 | 0 | 0 | 0 | 0 | 0 | 0 | 0 | 0 | 0 |
| NY48 testes | 0 | 0 | 0 | 0 | 1 | 0 | 0 | 0 | 0 | 0 | 0 | 0 | 1 |
| NY48 mal car | 0 | 1 | 0 | 2 | 1 | 0 | 0 | 0 | 0 | 1 | 0 | 0 | 3 |
| NY48 ovary | 0 | 0 | 0 | 0 | 0 | 0 | 0 | 0 | 0 | 0 | 0 | 0 | 0 |
| NY48 fem car | 0 | 0 | 0 | 0 | 0 | 0 | 0 | 0 | 0 | 0 | 0 | 0 | 0 |
| NY56 testes | 0 | 1 | 0 | 3 | 2 | 0 | 0 | 2 | 0 | 0 | 0 | 0 | 5 |
| NY56 mal car | 0 | 1 | 0 | 4 | 1 | 0 | 0 | 3 | 0 | 1 | 0 | 1 | 6 |
| NY56 ovary | 0 | 1 | 0 | 0 | 0 | 0 | 0 | 0 | 0 | 1 | 0 | 0 | 1 |
| NY56 fem car | 0 | 1 | 0 | 0 | 0 | 0 | 0 | 1 | 0 | 0 | 0 | 0 | 1 |
| NY81 testes | 0 | 2 | 0 | 1 | 3 | 0 | 0 | 0 | 0 | 1 | 0 | 0 | 5 |
| NY81 mal car | 0 | 2 | 0 | 1 | 2 | 0 | 0 | 1 | 0 | 0 | 0 | 0 | 4 |
| NY81 ovary | 0 | 0 | 0 | 0 | 0 | 0 | 0 | 0 | 0 | 0 | 0 | 0 | 0 |
| NY81 fem car | 0 | 0 | 0 | 0 | 0 | 0 | 0 | 0 | 0 | 0 | 0 | 0 | 0 |
| NY85 testes | 0 | 1 | 0 | 1 | 2 | 0 | 1 | 0 | 0 | 0 | 0 | 0 | 4 |
| NY85 mal car | 0 | 2 | 3 | 2 | 2 | 0 | 1 | 2 | 2 | 2 | 0 | 0 | 12 |
| NY85 ovary | 0 | 1 | 0 | 0 | 1 | 0 | 0 | 1 | 0 | 2 | 0 | 0 | 3 |
| NY85 fem car | 0 | 0 | 0 | 0 | 0 | 0 | 1 | 1 | 0 | 1 | 0 | 0 | 2 |
| CY22B testes | 0 | 2 | 3 | 1 | 2 | 0 | 1 | 2 | 0 | 0 | 0 | 1 | 9 |
| CY22B mal car | 0 | 3 | 0 | 3 | 4 | 0 | 1 | 3 | 0 | 0 | 0 | 1 | 10 |
| CY22B ovary | 0 | 0 | 2 | 0 | 1 | 0 | 0 | 0 | 0 | 0 | 0 | 0 | 3 |
| CY22B fem car | 0 | 1 | 0 | 1 | 0 | 0 | 0 | 0 | 0 | 0 | 0 | 0 | 1 |
| CY21B3 testes | 0 | 2 | 1 | 0 | 5 | 0 | 1 | 1 | 0 | 1 | 0 | 0 | 9 |
| CY21B3 mal car | 0 | 3 | 1 | 1 | 5 | 1 | 1 | 2 | 0 | 0 | 0 | 1 | 11 |
| CY21B3 ovary | 0 | 1 | 1 | 0 | 3 | 0 | 0 | 0 | 0 | 1 | 0 | 0 | 5 |
| CY21B3 fem car | 0 | 0 | 0 | 0 | 2 | 0 | 0 | 0 | 0 | 0 | 0 | 0 | 2 |
| CY20A testes | 0 | 2 | 2 | 1 | 4 | 1 | 0 | 1 | 0 | 3 | 0 | 0 | 10 |
| CY20A mal car | 0 | 1 | 1 | 1 | 2 | 1 | 0 | 2 | 0 | 3 | 0 | 2 | 8 |
| CY20A ovary | 0 | 1 | 0 | 0 | 3 | 1 | 0 | 0 | 0 | 2 | 0 | 0 | 5 |
| CY20A fem car | 0 | 0 | 0 | 0 | 3 | 1 | 0 | 0 | 0 | 1 | 0 | 0 | 4 |
| CY28A4 testes | 2 | 4 | 2 | 4 | 0 | 0 | 2 | 1 | 0 | 0 | 0 | 1 | 11 |
| CY28A4 mal car | 0 | 3 | 1 | 4 | 1 | 0 | 2 | 2 | 0 | 0 | 0 | 3 | 10 |
| CY28A4 ovary | 0 | 2 | 0 | 2 | 0 | 0 | 1 | 0 | 0 | 0 | 0 | 0 | 3 |
| CY28A4 fem car | 0 | 0 | 0 | 0 | 0 | 0 | 1 | 0 | 0 | 0 | 0 | 0 | 1 |
| CY04B testes | 0 | 3 | 2 | 1 | 3 | 0 | 0 | 3 | 1 | 1 | 0 | 2 | 11 |
| CY04B mal car | 0 | 3 | 0 | 1 | 2 | 0 | 0 | 5 | 0 | 0 | 0 | 3 | 8 |
| CY04B ovary | 0 | 1 | 0 | 0 | 2 | 0 | 0 | 0 | 0 | 1 | 0 | 0 | 3 |
| CY04B fem car | 0 | 1 | 0 | 0 | 0 | 0 | 0 | 3 | 0 | 0 | 0 | 2 | 3 |
| CY17C testes | 0 | 2 | 1 | 1 | 3 | 0 | 0 | 1 | 0 | 0 | 0 | 0 | 6 |
| CY17C mal car | 0 | 2 | 1 | 6 | 3 | 0 | 0 | 2 | 1 | 4 | 0 | 0 | 12 |
| CY17C ovary | 0 | 0 | 0 | 0 | 0 | 0 | 0 | 0 | 0 | 0 | 0 | 0 | 0 |
| CY17C fem car | 0 | 0 | 0 | 0 | 0 | 0 | 0 | 1 | 0 | 0 | 0 | 1 | 1 |
| CY08A testes | 0 | 2 | 0 | 2 | 3 | 0 | 0 | 0 | 1 | 0 | 0 | 0 | 6 |
| CY08A mal car | 0 | 3 | 1 | 3 | 3 | 0 | 0 | 2 | 0 | 1 | 0 | 1 | 9 |
| CY08A ovary | 0 | 1 | 1 | 0 | 1 | 0 | 0 | 0 | 0 | 1 | 0 | 0 | 3 |
| CY08A fem car | 0 | 0 | 1 | 0 | 0 | 0 | 0 | 0 | 0 | 0 | 0 | 0 | 1 |

**S2 Table:** Number of genes within 1kb of rearrangement calls that are either down regulated or up regulated in each individual tissue.

| Tissue | Up regulated | Down Regulated | Both* |
|---|---|---|---|
| **Testes** | 18 | 21 | 2 |
| **Male Carcass** | 15 | 30 | 6 |
| **Ovaries** | 16 | 32 | 2 |
| **Female Carcass** | 18 | 18 | 0 |

* Both indicates that in some lines that have a certain rearrangement near the gene has up regulated expression while in other lines that have the rearrangement the gene is down regulated.

**S3 Table:** Tukey multiple comparisons of means of total rearrangement sites found per base pair on each major chromosome arm

| | Differential | Lower end point | Upper end point | p adj |
|---|---|---|---|---|
| 2R-2L | 1.15E-07 | -5.55E-07 | 7.86E-07 | 0.9883147 |
| 3L-2L | -2.85E-07 | -9.56E-07 | 3.86E-07 | 0.7509075 |
| 3R-2L | -8.73E-07 | -1.54E-06 | -2.02E-07 | 0.0048835 |
| X-2L | 2.06E-06 | 1.39E-06 | 2.73E-06 | 0 |
| 3L-2R | -4.00E-07 | -1.07E-06 | 2.70E-07 | 0.4507802 |
| 3R-2R | -9.88E-07 | -1.66E-06 | -3.17E-07 | 0.0010817 |
| X-2R | 1.94E-06 | 1.27E-06 | 2.62E-06 | 0 |
| 3R-3L | -5.88E-07 | -1.26E-06 | 8.31E-08 | 0.1120626 |
| X-3L | 2.35E-06 | 1.67E-06 | 3.02E-06 | 0 |
| X-3R | 2.93E-06 | 2.26E-06 | 3.60E-06 | 0 |

**S4 Table:** Pairwise comparisons between chromosome arms in abundance of rearrangements within chromosomes

| | Differential | Lower end point | Upper end point | p adj |
|---|---|---|---|---|
| 2R-2L | 5.06E-09 | -2.37E-07 | 2.47E-07 | 0.9999971 |
| 3L-2L | -7.15E-09 | -2.49E-07 | 2.35E-07 | 0.9999886 |
| 3R-2L | -2.12E-07 | -4.54E-07 | 2.99E-08 | 0.1116779 |
| X-2L | 8.00E-07 | 5.58E-07 | 1.04E-06 | 0 |
| 3L-2R | -1.22E-08 | -2.54E-07 | 2.30E-07 | 0.9999037 |
| 3R-2R | -2.17E-07 | -4.59E-07 | 2.48E-08 | 0.0981546 |
| X-2R | 7.95E-07 | 5.53E-07 | 1.04E-06 | 0 |
| 3R-3L | -2.05E-07 | -4.47E-07 | 3.70E-08 | 0.1332882 |
| X-3L | 8.08E-07 | 5.66E-07 | 1.05E-06 | 0 |
| X-3R | 1.01E-06 | 7.71E-07 | 1.25E-06 | 0 |

**S5 Table:** pairwise comparisons between chromosome arms in abundance of rearrangements between chromosomes

| | Differential | Lower end point | Upper end point | p adj |
|---|---|---|---|---|
| 2R-2L | 1.05E-07 | -2.66E-07 | 4.76E-07 | 0.9287468 |
| 3L-2L | -2.71E-07 | -6.42E-07 | 1.00E-07 | 0.2520333 |
| 3R-2L | -4.48E-07 | -8.19E-07 | -7.73E-08 | 0.0104721 |
| X-2L | 4.59E-07 | 8.83E-08 | 8.30E-07 | 0.008239 |
| 3L-2R | -3.76E-07 | -7.47E-07 | -5.06E-09 | 0.0454781 |
| 3R-2R | -5.54E-07 | -9.24E-07 | -1.83E-07 | 0.0009076 |
| X-2R | 3.54E-07 | -1.70E-08 | 7.25E-07 | 0.0682095 |
| 3R-3L | -1.78E-07 | -5.48E-07 | 1.93E-07 | 0.659954 |
| X-3L | 7.30E-07 | 3.59E-07 | 1.10E-06 | 0.0000091 |
| X-3R | 9.07E-07 | 5.37E-07 | 1.28E-06 | 0.0000001 |

**S6 Table:** Total number of rearrangement sites that have coverage depth twice the average coverage depth of each line. Coverage depth of each rearrangement plus 100 base pair regions or 500 base pair regions flanking each rearrangement site were calculated and compared to the average coverage depth. The number of sites that had double the coverage depth divided by the total number of rearrangement sites for that line is represented in the percent columns.

| | Average Coverage depth | Total sites | 500 bp | 500 bp percent | 100 bp | 100 bp percent |
|---|---|---|---|---|---|---|
| *NY48* | 34.5 | 430 | 33 | 7.6 | 91 | 21.2 |
| *NY56** | 12.1 | 192 | 69 | 35.9 | 131 | 68.2 |
| *NY62* | 44.8 | 502 | 33 | 6.6 | 80 | 15.9 |
| *NY66* | 26.5 | 316 | 33 | 8.9 | 71 | 22.5 |
| *NY73* | 27.6 | 372 | 39 | 10.5 | 91 | 24.5 |
| *NY81* | 23.9 | 332 | 24 | 7.2 | 61 | 18.4 |
| *NY85* | 61.1 | 592 | 19 | 3.2 | 48 | 8.1 |
| *CY08A* | 37.5 | 520 | 21 | 4 | 37 | 7.1 |
| *CY20A* | 93.7 | 846 | 21 | 2.5 | 88 | 10.4 |
| *CY28A4* | 58.3 | 652 | 25 | 3.8 | 85 | 13 |
| *CY04B* | 64.3 | 910 | 25 | 2.7 | 84 | 9.2 |
| *CY22B* | 45.5 | 414 | 8 | 1.9 | 33 | 8 |
| *CY21B3* | 44.8 | 458 | 6 | 1.3 | 36 | 7.9 |
| *CY17C* | 43.2 | 488 | 13 | 2.7 | 34 | 7 |

*with average depth of only 12.1, the total number of rearrangements are more likely an over representation than the total number of possible duplications.

**S7 Table:** Number and ratios in which the abnormal read pairs were aligned in the same direction. The reads lining in the same direction indicates either the translocation was inserted inverted or (if within chromosome) the rearrangement identified is an inversion.

| Line | Coverage depth | Total Within chromosomes | Within chromosomes linear reads | Within chromosomes Percent | Total Between chromosomes | Between chromosomes linear reads | Between chromosome Percent |
|---|---|---|---|---|---|---|---|
| *NY73* | 27.6 | 57 | 31 | 54% | 129 | 62 | 48% |
| *NY66* | 26.5 | 47 | 22 | 47% | 111 | 60 | 54% |
| *NY62* | 44.8 | 74 | 39 | 53% | 177 | 96 | 54% |
| *NY48* | 34.5 | 59 | 35 | 59% | 156 | 66 | 42% |
| *NY56* | 12.1 | 26 | 14 | 54% | 70 | 33 | 47% |
| *NY81* | 23.9 | 47 | 28 | 60% | 119 | 55 | 46% |
| *NY85* | 61.1 | 82 | 46 | 56% | 214 | 97 | 45% |
| *CY22B* | 45.5 | 60 | 24 | 40% | 148 | 62 | 42% |
| *CY21B3* | 44.8 | 74 | 35 | 47% | 155 | 76 | 49% |
| *CY20A* | 93.7 | 102 | 48 | 47% | 321 | 155 | 48% |
| *CY28A4* | 58.3 | 105 | 52 | 50% | 221 | 114 | 52% |
| *CY04B* | 64.3 | 129 | 73 | 57% | 326 | 150 | 46% |
| *CY17C* | 43.2 | 73 | 33 | 45% | 171 | 82 | 48% |
| *CY08A* | 37.5 | 70 | 36 | 51% | 190 | 81 | 43% |

**S8 Table:** GO terms all genes with *D. melanogaster* ortholog (733) associated with rearrangements

| Cluster | count | enrichment | P_value | Benjamini |
|---|---|---|---|---|
| Alternative Splicing | 82 | 8.39 | 4.30E-14 | 1.00E-11 |
| Membrane | 240 | 7.18 | 1.60E-09 | 1.90E-07 |
| Glycoprotein | 53 | 3.56 | 2.20E-06 | 7.70E-05 |
| Pleckstrin | 22 | 3.21 | 6.00E-05 | 2.10E-02 |
| SH3 homology/domain | 14 | 2.7 | 1.80E-04 | 3.20E-02 |
| protein kinase/transferase | 38 | 2.4 | 1.00E-07 | 3.60E-05 |
| dorsal closure head/eye development | 16 | 2.93 | 4.00E-04 | 3.80E-02 |
| PDZ domain | 13 | 2. 06 | 1.80E-04 | 1.80E-012 |

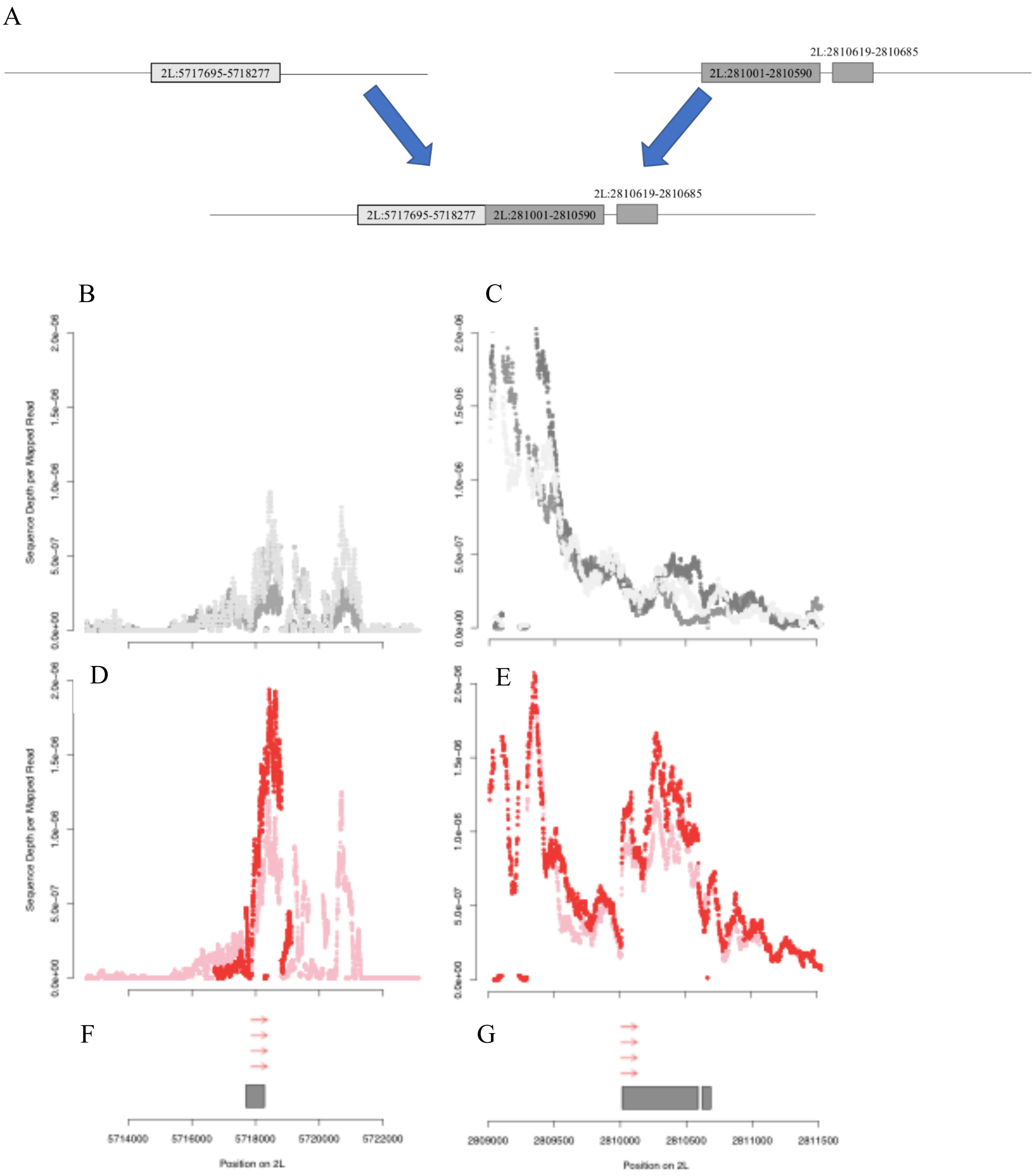

**S1 Figure:** New gene formation through genome rearrangement on chromosome 3R and 2L. The gene (*GE18374*)/regulatory elements at 2L:5171796-5718110 appears to be altering expression at 2L:2810011-2810533. A) Diagram showing the predicted sequence movement based on the Trinity Transcript blast. Observing sequence depth of the RNA we can infer relative expression and identify newly transcribed regions in lines that have rearrangement calls. B) and C) The grey coverage lines are RNA sequence coverage from Tophat of three replicates of the reference line male carcass and 3 replicates of reference line testes which do not have this rearrangement. D) and E) CY17C male carcass (red line), testes (pink line) has a rearrangement between 2L:5171796-5718110 and 2L:2810011-2810533. F) and G) Red arrows indicate the genomic rearrangement calls, both reads point in the same direction that suggest the transported section of DNA was inserted inversely. The grey boxes represent the Trinity transcript mapped to the reference genome.

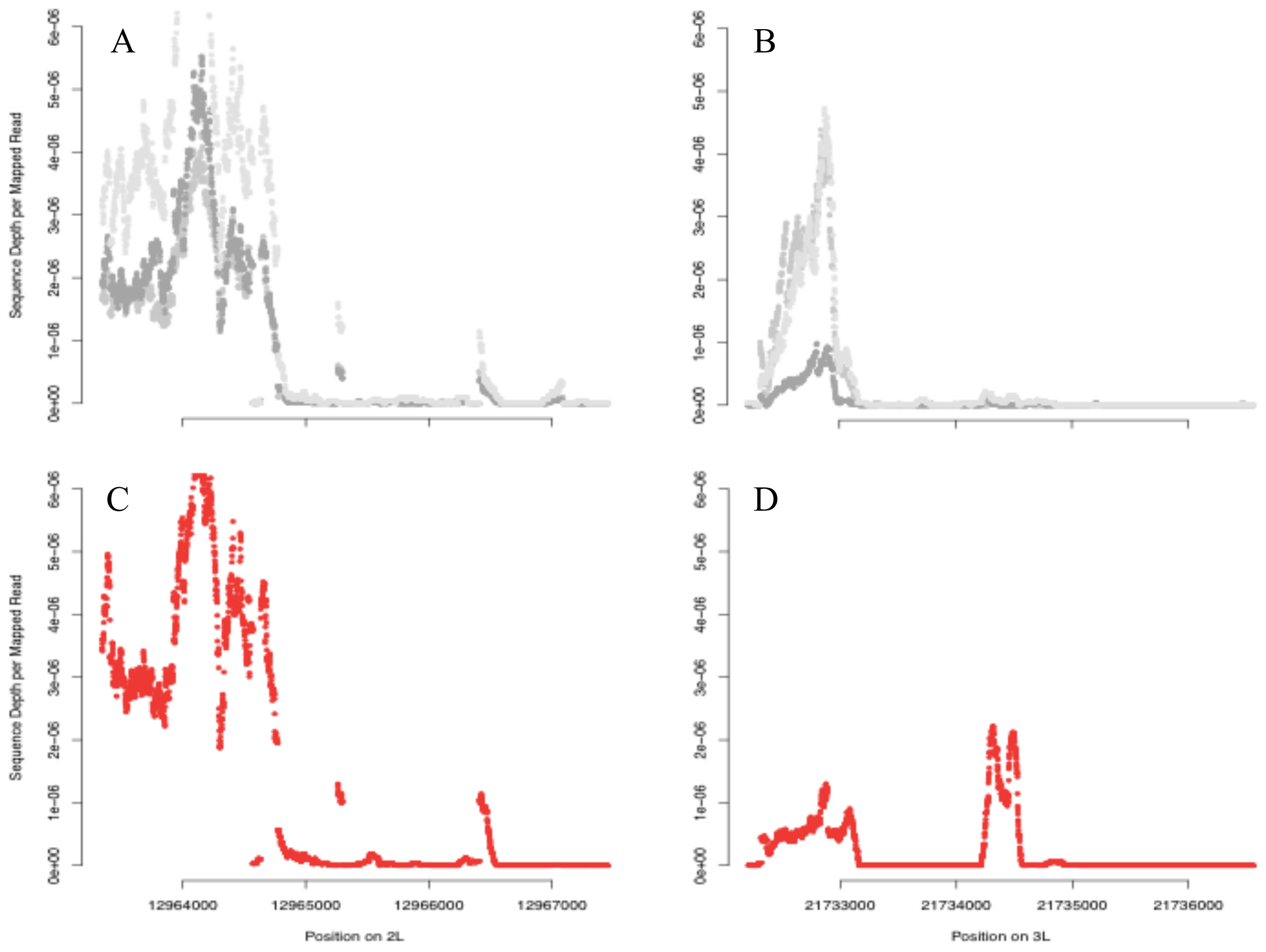

**S2 Figure:** A) and B) The grey coverage lines are RNA sequence coverage from Tophat of 3 replicates of reference line testes which do not have this rearrangement. C) and D) CY22B testes has a rearrangement between 2L:12965344-12965453 and 3L:21734209-21734559. This rearrangement appears to have created a newly transcribed region 3L:21734209-21734559. This may have been altered by the gene *GE26196* or its surrounding regulatory elements near 2L:12965344-12965453. However, this rearrangement only has 3 supporting Illumina genomic reads. A 4$^{th}$ read pair can be found 25bp over our 325bp coupling limit. This suggest that our conservative approach to the identification of new genes is an underestimate of the overall new genes created by rearrangements.

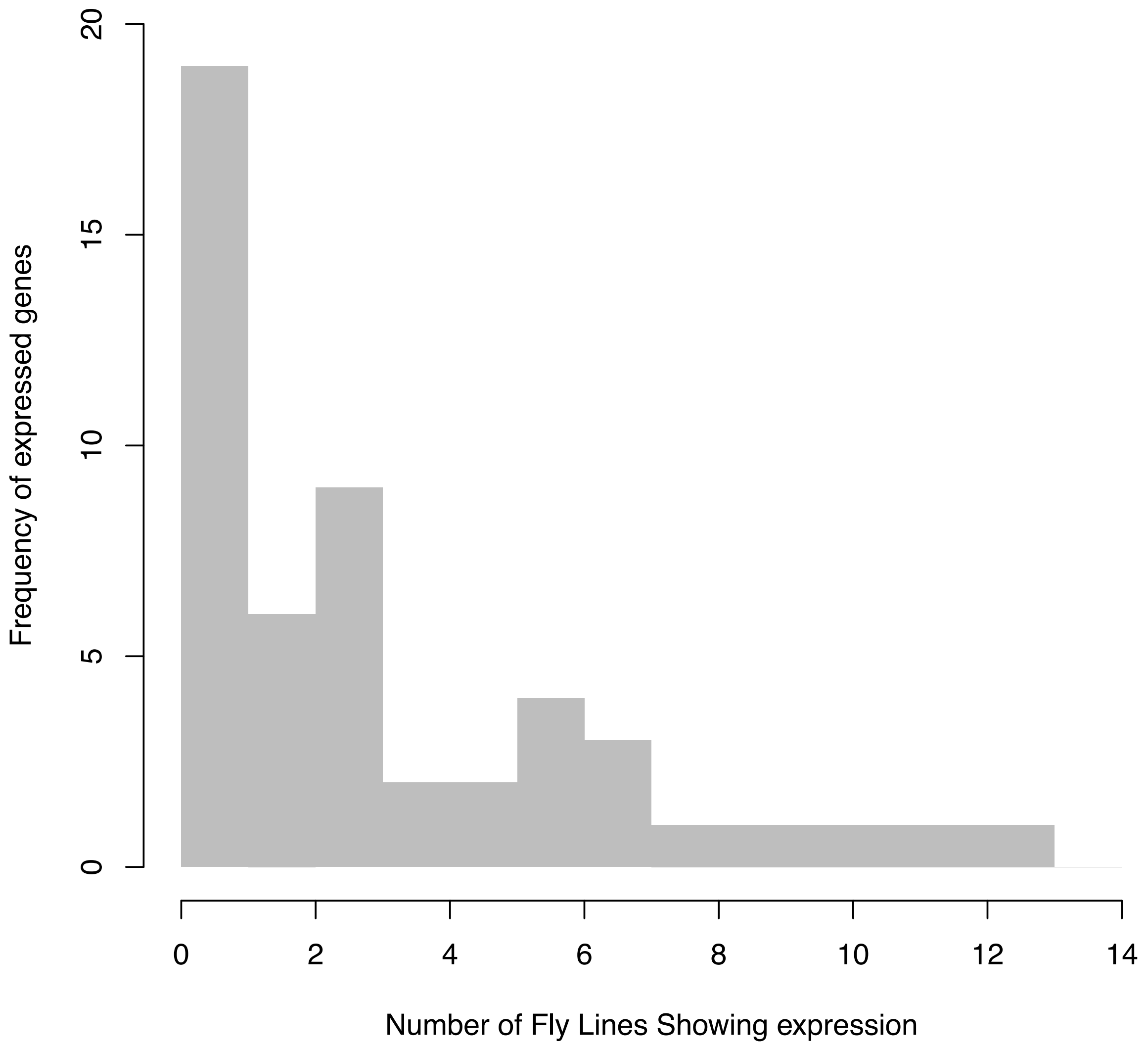

**S3 Figure:** Site frequency spectrum of rearrangements that are associated with fusion transcripts found in the 14 lines. Most new genes formed at rearrangements are low-frequency variation.

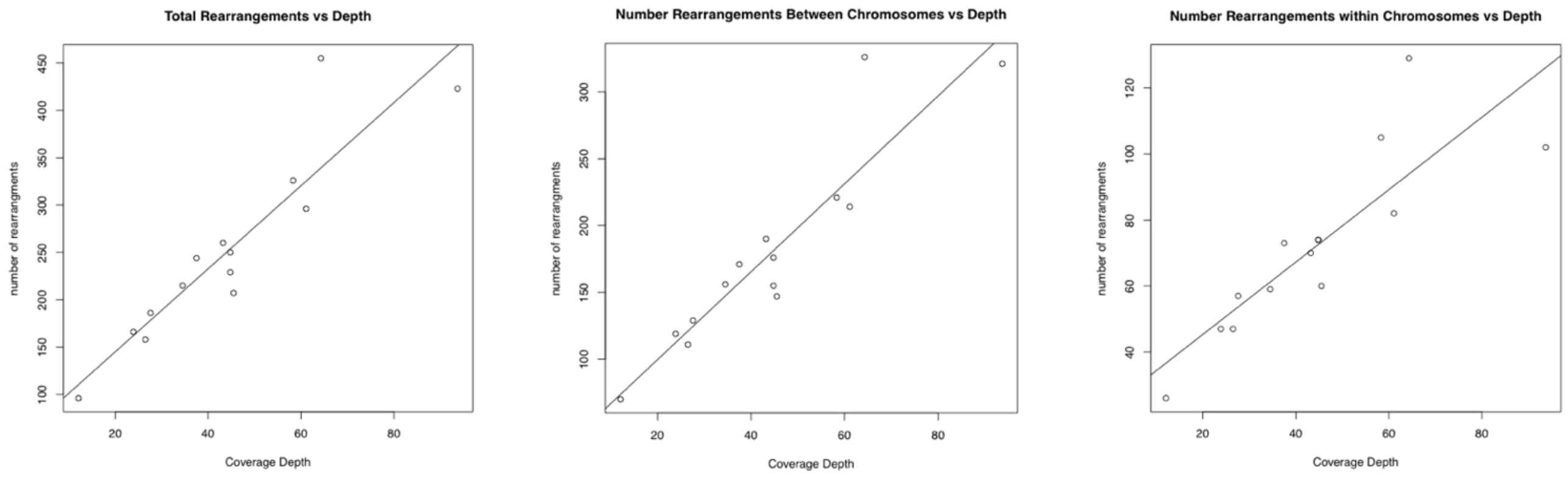

**S4 Figure:** A) Total, B) between, C) and within chromosome rearrangements identified that have 4 supporting independent read-pairs. There is a strong correlation between sequence coverage depth and total number of rearrangements ($R^2$=0.8231, $P<4.7\mathrm{x}10^{-6}$), between chromosomes rearrangements ($R^2$ =0.8443, p<2.2e$^{-06}$), and within chromosome rearrangements ($R^2$ =0.6936, $P<1.4\mathrm{x}10^{-4}$).

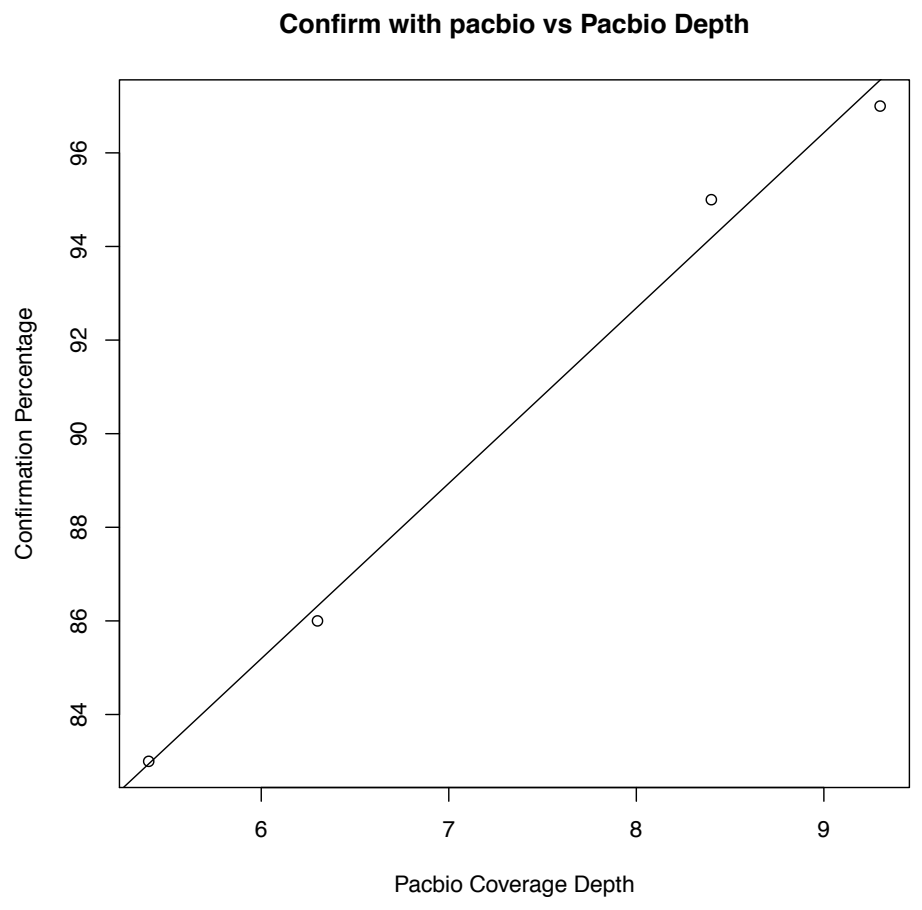

**S5 Figure:** Percent of the rearrangements identified by illumine paired-end sequencing confirmed by PacBio sequencing over the coverage depth of the PacBio sequencing We aligned PacBio sequence reads to the *D. yakuba* reference using a BLASTn with the repetitive DNA filter turned off and an e-value cutoff of $10^{-10}$. If a single molecule read blast within 2kb of the genomic rearrangement call it was counted as confirmation.

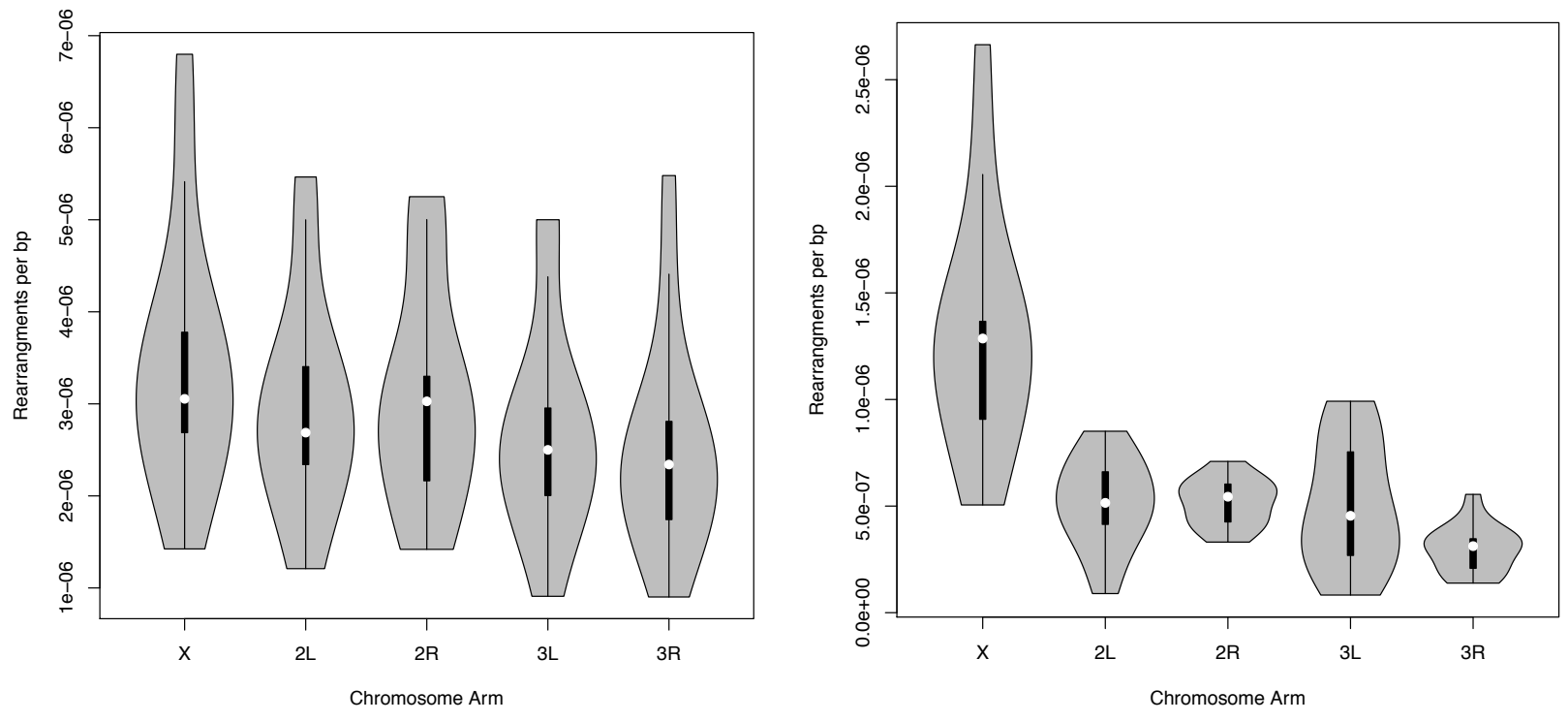

**S6 Figure:** Number of chromosomal rearrangement breakpoints per base pair on each chromosome arm for each *D. yakuba* line. A) Number of rearrangement brekapoints between chromosomes each rearrangement was counted for both arms involved. Chromosome rearrangements, the X chromosome shows significantly higher number of within chromosome rearrangements with 261 of the 671 total rearrangements (ANOVA, $F(4, 52)= 42.06$, $P<10^{-14}$, Tukey HSD for each X comparison $P<10^{-7}$ ). While none of the autosomes are significantly different with respect to number of rearrangements per base pair (Tukey HSD for all comparisons, $P>0.05$). B) Number of rearrangement breakpoints with both ends on the same chomosome arm and 1Mb distally between the two ends varied by chromosome arm (ANOVA, $F(4, 52)= 14.26$, $P<10^{-7}$) We found that the X chromosome had significantly more rearrangement sites than chromosome arms 2L (Tukey HSD, $P<0.009$), 3L ( Tukey HSD, $P<10^{-5}$), and 3R (Tukey, HSD $P<10^{-6}$) but not 2R (Tukey HSD, $P=0.068$). Chromosome arm 2R showed greater number of between chromosome rearrangements per base pair than chromosome arms 3L (Tukey HSD, $P<0.05$), and 3R (Tukey HSD, $P<0.001$).

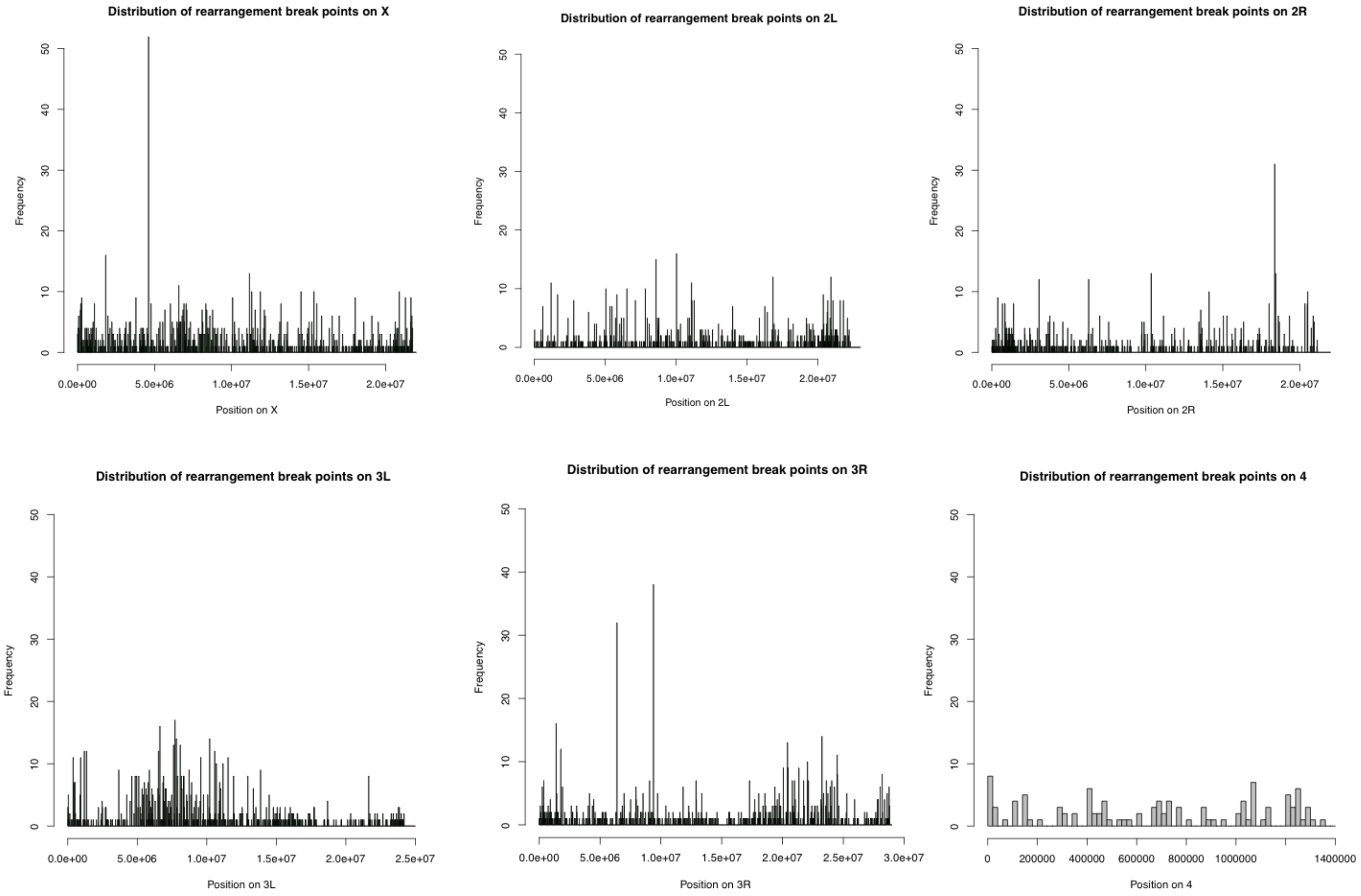

**S7 Figure:** Distribution of rearrangement sites along the 4 chromosome arms using 20kb windows. There are 4 rearrangement hotspots, one on X, one on 2R, and two on 3R.

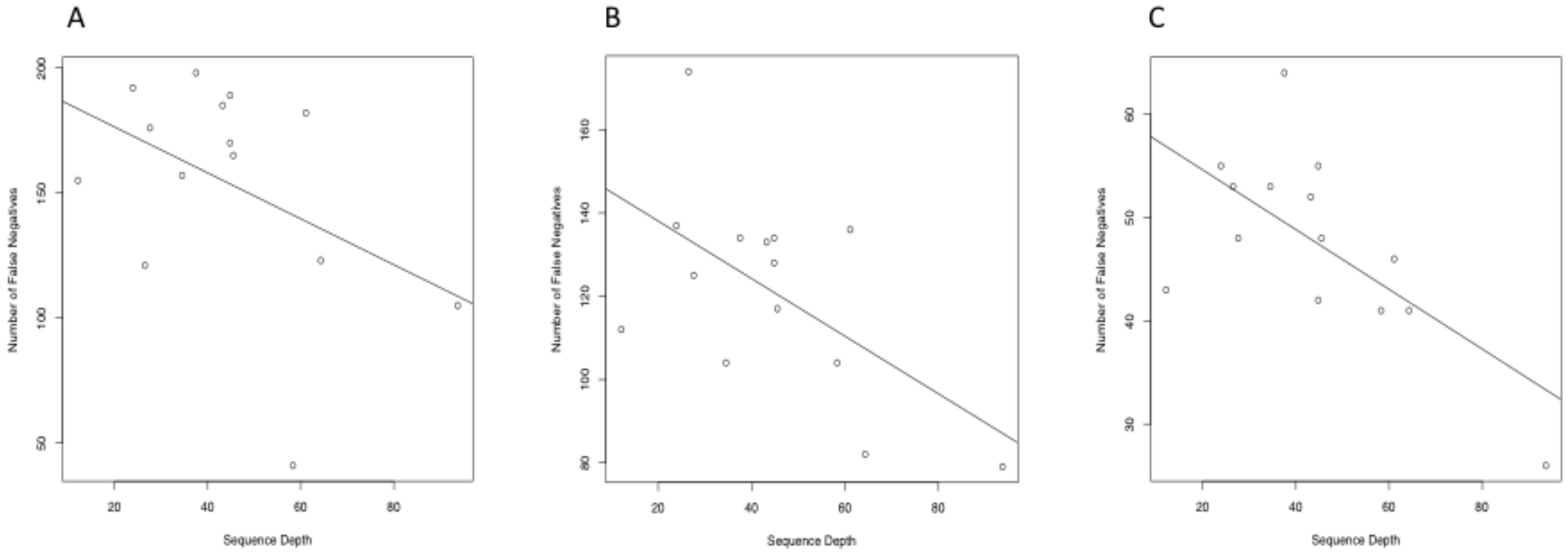

**S8 Figure:** A) Total Number of possible false negatives of each line against sequence coverage depth of that line. B) Number of possible false negative between chromosomes, C) number of false negatives within chromosome rearrangements. There is not a significant correlation between the total number of false negatives and coverage depth (*$R^2$*= 0.1189, *P*=0.1189), however isolating between chromosome and within chromosome rearrangements they both show a negative correlation between coverage depth and false negative estimates (between chromosomes *$R^2$*=0.2839, *P*=0.02892; within chromosomes *$R^2$*=.3841, *P*=0.0107). Possible false negatives were classified by having 1-3 read pair support for a rearrangement that is found in at least one other line with 4 or greater read pair support.

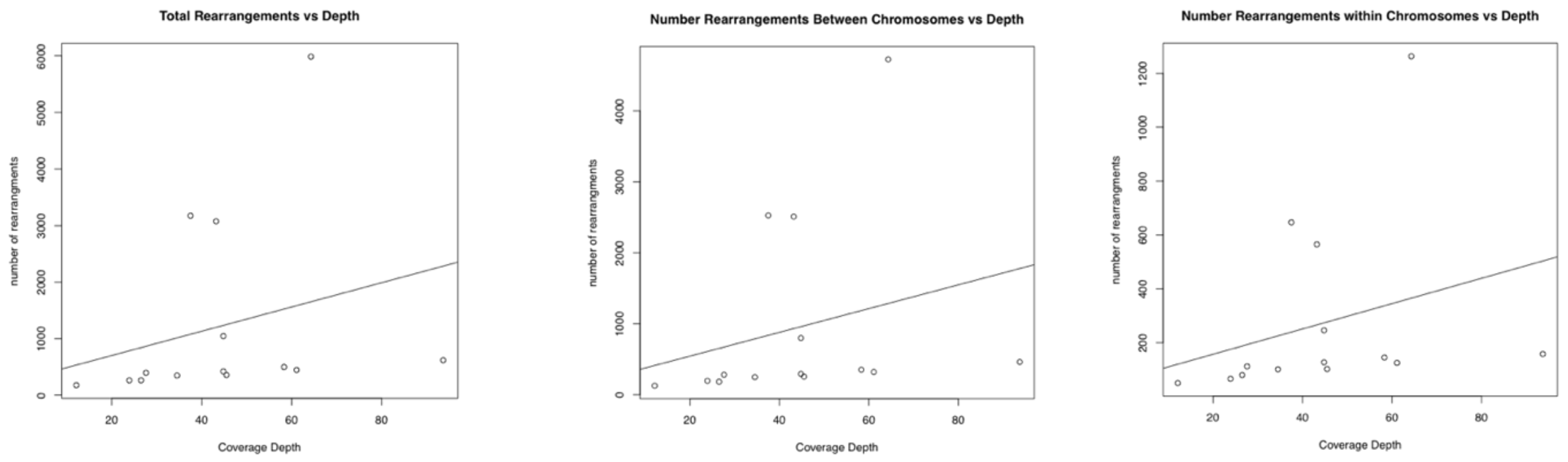

**S9 Figure:** A) Total, B) between, C) and within chromosome rearrangements identified that have 3 supporting independent read-pairs. There is no correlation between sequence coverage depth and total number of rearrangements ($R^2$=0.009632, P=0.3678), between chromosomes rearrangements ($R^2$ =0.0132, P=0.38), and within chromosome rearrangements ($R^2$ =0.004914, P=0.3226). Lines CY04B, CY08A, CY17C, and CY21B3 have abnormally high amount of structure calls, most are rearrangements that have support of three read-pairs.